\documentclass[aps,twocolumn,floats,pre,amsmath,superscriptaddress]{revtex4}
\usepackage{graphics,graphicx,epsfig}
\usepackage{amssymb,color}
\usepackage{epsf,epstopdf,wrapfig}

\newcommand{\beq}{\begin{equation}}
\newcommand{\eeq}{\end{equation}}
\newcommand{\beqn}{\begin{eqnarray}}
\newcommand{\eeqn}{\end{eqnarray}}

 \let\b=\beta  \let\d=\delta
   \let\k=\kappa
\let\l=\lambda    
\let\s=\sigma \let\t=\tau  
   
\let\D=\Delta   
\let\Si=\Sigma   
   \let\io=\infty

\def\PP{{\cal P}} 
 
\def\TT{{\cal T}}\def\NN{{\cal N}} 
\def\LL{{\cal L}}

\def\to{\rightarrow} \def\la{\langle} \def\ra{\rangle}

\def\(({\left(}
\def\)){\right)}                       
\def\[[{\left[}
\def\]]{\right]}

\newcommand{\bra}[1]{\langle{#1}|}
\newcommand{\ket}[1]{|{#1}\rangle{}}

\newcommand{\bea}{\begin{eqnarray}}
\newcommand{\eea}{\end{eqnarray}}

\newcommand{\bs}{ \mbox{\boldmath$\sigma$}}

\begin{document}

\title{Transition path sampling algorithm for discrete many-body systems}

\author{Thierry Mora}
\affiliation{Laboratoire de Physique Statistique, UMR 8550, CNRS and Ecole Normale Sup\'erieure, 
24 Rue Lhomond, 75231 Paris Cedex 05, France}

\author{Aleksandra M. Walczak}
\affiliation{Laboratoire de Physique Th\'eorique, UMR 8549, CNRS and Ecole Normale Sup\'erieure, 
24 Rue Lhomond, 75231 Paris Cedex 05, France}

\author{Francesco Zamponi}
\affiliation{Laboratoire de Physique Th\'eorique, UMR 8549, CNRS and Ecole Normale Sup\'erieure, 
24 Rue Lhomond, 75231 Paris Cedex 05, France}

\date{\today}

\begin{abstract}
We propose a new Monte Carlo method for efficiently sampling trajectories with fixed initial and final conditions in a system with discrete degrees of freedom.
The method can be applied to any stochastic process with local interactions, including systems that are out of equilibrium. We combine the proposed path-sampling algorithm with thermodynamic integration to calculate transition rates. We demonstrate our method on the well studied 2D Ising model with periodic boundary conditions, and show agreement with other results both for large and small system sizes. 
The method scales well with the system size, allowing one to simulate systems with many degrees of freedom, and providing complementary information with 
respect to other algorithms.
\end{abstract}

\maketitle

\section{Introduction}

A common feature of complex systems is the existence of local attractors separated by high activation barriers \cite{hangii_review, valerianirev}. 
When considering the dynamics on such landscapes, one often finds the system trapped in these metastable states. 
The long-term dynamics in these systems is then dominated by long periods of local equilibration inside the metastable states, separated by rare jumps from one state to another. 
The simplest example is a continuous degree of freedom moving in a potential with only two minima, 
which correspond to two peaks of its steady-state probability distribution, separated by an energy barrier.
This problem can be tackled analytically, and in some cases more complex problems can be mapped on it
by defining a one-dimensional reaction coordinate along which the 
transition rates between the two metastable states can be calculated. However, in most real systems, even those with few degrees of freedom, 
the definition of a unique reaction coordinate is often not possible, 
and one must attempt to sample the reaction or transition paths from one metastable state to another exhaustively. 
Yet the rarity of these transition events makes usual simulation techniques, 
which are based on sampling of all possible trajectories, incredibly time consuming. 
Naturally, the difficulty of sampling grows with the number of degrees of freedom of the system. 
Many efficient algorithms have been developed to calculate transition rates efficiently, 
but often these techniques \cite{bennet, dinner} are limited to systems obeying detailed 
balance since they require knowing the phase space density. 
Recently a number of methods applicable to nonequilibrium systems have been 
developed~\cite{valerianirev, chandler, dellago2, tenwoldeffs, weinane_erik, erik2, kurchan1, kurchan2, picciani}, which effectively calculate 
the flux of probability between the steady states. In this paper we present a new Monte Carlo 
technique for sampling transition paths with fixed initial and final
conditions in nonequilibrium systems. The technique
adapts the transition path sampling~\cite{chandler, dellago2} method to discrete systems, 
and is based on the local update of single-variable paths~\cite{francescoprb}. 
We show how this new method allows us to calculate transition rates.

Metastable states appear in many natural systems, and the problem of transitions from these states has been extensively studied.
For example, in magnetic systems below the critical temperature, the system gets trapped in one of its low-energy spin configurations 
and rarely explores the intermediate states in between. In frustrated spin systems, the number of possible metastable 
states increases rapidly with the system size, leading to a very rugged landscape. On the contrary, 
in ferromagnetic systems there are typically two low energy states---all spins up and all spins down.
The simplest example is the mean-field formulation of the Ising ferromagnet (the so-called Curie-Weiss model), where 
transition rates can be found exactly~\cite{langer} by reducing the problem to one dimension. 
Another well-studied example is the two-dimensional Ising model, 
where the asymptotic large-size scaling for the transition rate has been calculated rigorously thanks to a detailed understanding
of the thermodynamics \cite{onsager, binder, gallavotti} and of the dynamics of the model~\cite{martinelli}.

A lot of progress in the development of methods aimed at calculating transition rates between metastable states has been made 
in the context of chemical reactions \cite{hangii_review, chandler, wolynes}. One of these specific methods, 
which requires no prior knowledge of the transition states, relies on the statistical sampling of paths by means
of a Monte Carlo simulation on the paths themselves, which are treated as the microscopic states of the system 
and whose action plays the role of an energy \cite{chandler}:
paths are therefore sampled according to their action.
The resulting method is a finite-temperature generalization 
of the eikonal or WKB method in which one finds the most probable (lowest action) path, around which the contribution 
of all transition paths in calculated within a quadratic approximation. 

The string method~\cite{weinane_erik} instead identifies the trajectories which carry most of the probability current, 
by constructing a system of interfaces between the two metastable states in a deterministic way.
Other methods have considered the flux of probability between states by constructing a system of interfaces or benchmarks, 
and sampling trajectories between them with a genetic algorithm (only survive the paths that pass the benchmarks) 
to estimate the probability of survival across all interfaces, from which the transition rate is calculated~\cite{dellago2, valerianirev}.
In a similar spirit, 
cloning techniques have been used to select, in a population of random walkers, 
those that correctly sample the transition path~\cite{kurchan1,kurchan2,picciani}.

It is worth mentioning that many of these nonequilibrium methods have been developed with biological systems in mind. For example, gene expression in cells can often lead to the formation of a multistable systems, corresponding to different expression levels of proteins which have been associated with cell types \cite{kauffman}. Other applications include membrane pore formation and conformation changes in polymers \cite{valerianirev, erik2}.

We generalized the transition path sampling technique of~\cite{chandler} to discrete many-body systems, 
and obtained an algorithm that should allow for the effective calculation of transition rates between the metastable states of complex systems. 
The method does not require the explicit forward Monte Carlo simulation of the system, but instead performs a Monte Carlo search directly on the paths, 
under the constraint of fixed initial and final conditions. The method is quite general and can be applied to a number of systems. 
It is most effective when the system comprises many variables transitioning between discrete states, 
and when the dependence of each variable on the rest of the system only involves a small subset of the other variables, 
or said differently, when the graph representing interactions between variables is sparse.

Throughout this paper, for ease of presentation, we describe our method on the example of a ferromagnetic spin system, 
but the method easily generalizes to any out-of-equilibrium system with discrete variables. 
The method can be briefly described as follows. 
It consists of a Monte Carlo Markov chain on spin trajectories, where each move involves the update of the path of a one spin at a time.
Consider a trajectory, or path, of many interacting spins over a given duration, with fixed initial and final conditions. 
The algorithm isolates the trajectory of a single spin chosen at random, leaving the paths of all other spins fixed or ``frozen''. 
It then generates at random a new path for this one spin, with a probability prescribed by the value of the other spins with which 
it interacts, and with constraints on its initial and final values.
This conditional sampling of a new single-spin path is performed using a transfer matrix technique across time. 
The procedure is repeated many times until the system of paths equilibrates, just like in a standard Monte Carlo dynamics, 
with the difference that here paths play the role of configurations.
We combine this sampling method with the technique of thermodynamic integration to calculate transition rates in the two-dimensional Ising ferromagnet. 

Our method shares similarities with the method of Dellago et al. \cite{chandler} and we discuss these similarities, as well as crucial differences, in the Conclusions. 
Our method can also be viewed as an application to stochastic systems of ideas presented in Krzakala {\em et al.} \cite{francescoprb} in the context of quantum spins.

\begin{figure}
\includegraphics[width =  \columnwidth]{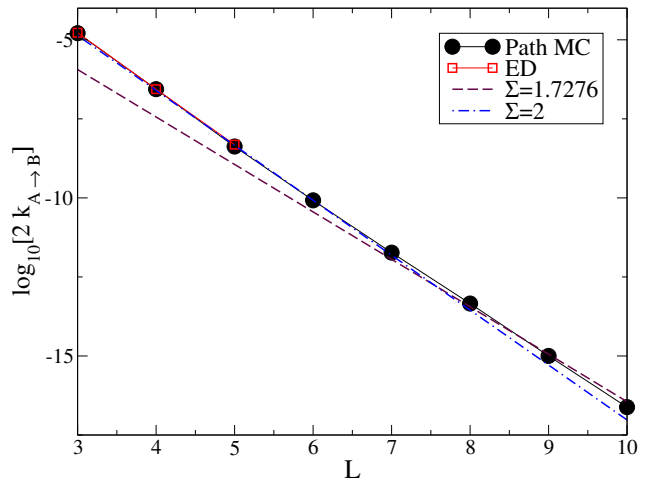}
\caption{A summary plot of the main results of this paper presented on
  the example of a 2D Ising ferromagnet with periodic boundary
  conditions. The plot represents the energy gap between the steady
  and first excited states, $\Delta$, as a function of the linear size
  of the system
  $L=\sqrt{N}$. The energy gap is equal to twice the transition rate
  between the two macroscopic states of the system (up and down).
  The results obtained from the path Monte Carlo sampling method presented in this paper are shown as full black circles;
  for small sizes, we also report results obtained from
  exact diagonalization of the master equation (open red squares). Dashed and dot-dashed lines represent asymptotic
  scalings (see section \ref{Asymscaling}).
  }
\label{FigureMAIN}
\end{figure}

After defining the problem we are setting out to study (Section \ref{sec:setup}), we recall some known results, in some cases providing a more compact derivation, 
on transition rates in the exactly solvable mean-field Ising model (Section \ref{sec:MF}). 
This simpler case will help us build some intuition for subsequent results. In Section \ref{sec:method}, we describe the method in the context of the 2D Ising model.
We then state the main results of this paper in Section \ref{sec:results}. These are best
summarized in Figure \ref{FigureMAIN}, where we show a perfect agreement between our method and exact matrix diagonalization of the master equation for small systems. 
Section \ref{sec:conclu} contains our conclusions.

\section{Definition of the problem}
\label{sec:setup}

In this section we give our basic definitions and notations about the class of models we study in this paper.
We introduce the Ising spin model with Glauber dynamics, and we write
the explicit Master equation describing its evolution.
Recall that, although we choose this specific setting to illustrate our method, the latter can be applied in a much more general setting,
namely for generic discrete systems undergoing a Markovian dynamics. In particular, the detailed balance condition is not required.

\subsection{Dynamics of an Ising spin system}
\label{Notation}

Consider a system of $N$ spins, interacting with each other via the Ising Hamiltonian:
\beq
H=-\sum_{i=1}^N \tilde{h}_i \sigma_i -\sum_{i,j} J_{ij} \sigma_i \sigma_j \ .
\label{EqHam}
\eeq
Let us denote a given spin configuration by $\bs = \{\sigma_i\}$. 
Under the assumption that the dynamics is Markovian in continuous time, it is entirely characterized by the instantaneous Poisson rates $w_{\bs;\bs'}$ of jumping from $\bs'$ to $\bs$.
The Master equation describing the evolution of the probability distribution of spin configurations, $p_t(\bs)$  can then be written as:
\bea
\partial_t p_t(\bs) &=& \sum_{\bs'} \left[ w_{\bs;\bs'} p_t(\bs')
- w_{\bs';\bs} p_t(\bs) \right]\\
\partial_t p_t &=& {\cal L} \, p_t,
\eea
where the evolution operator $\cal L$ is defined as ${\cal L}(\bs;\bs')=w_{\bs;\bs'}-\delta_{\bs,\bs'}\sum_{\bs''}w_{\bs'';\bs}$.
We specialize to dynamics where only one spin may flip at a time.
We denote by $\bs_{\backslash i}$ the set of ``all spins but $i$", and we denote by
$\bs_{\updownarrow i} = \{\bs_{\backslash i} , -\sigma_i \}$ the configuration that differs from
$\bs$ by a flip of spin $i$.
The variation of the Hamiltonian under one spin flip is
\beq
\Delta E = H(\bs) - H(\bs_{\updownarrow i}) = - 2 h_i \s_i \ ,
\eeq
with
\beq
h_i = \tilde{h}_i + \sum_{j (\neq i)} J_{ij} \s_j \ .
\eeq

We assume that $w_{\bs;\bs'}$ vanishes unless $\bs' = \bs_{\updownarrow i}$ for some $i$. Transition rates are assumed to only depend on the energy difference between the initial and final states.
In this case one has $w_{\bs;\bs_{\updownarrow i}} = w(\Delta E) = w(-2 h_i \sigma_i)$.

Therefore we can write the master equation as
\beq
\partial_t p_t(\bs) =
\sum_i \left[ w(-2 h_i \sigma_i) p_t(\bs_{\updownarrow i}) - w(2 h_i\sigma_i) p_t(\bs)\right]
\label{mastereqn}
\eeq
The first term describes the probability of flipping spin $i$, 
so that the system comes into the state $\bs$ from $\bs_{\updownarrow i}$. 
The rate $w(\Delta E) = w(-2 h_i \sigma_i)$ is the rate of flipping spin $i$ 
from $-\sigma_i$ to $\sigma_i$, which depends on the value 
of the effective external field $h_i$. 
The second term is just a normalization condition accounting for all events where
the system leaves $\bs$.

There are many ways to define $w(\Delta E)$ so that it is consistent with the detailed balance condition:
\beq
w(\Delta E) = e^{-\b \Delta E} w(-\Delta E) \ .
\eeq
Here we choose:
\beq\label{wrate}
w(\Delta E)= e^{-\b \Delta E/2} \ .
\eeq
Note that changing the overall normalization of the rates just amounts to a rescaling
of time. We stress once again that we choose these rates for convenience, but 
our method applies to any choice of rates, even if they do not satisfy detailed balance.

\subsection{Transitions between two states}

\begin{figure}
\includegraphics[width =  \linewidth]{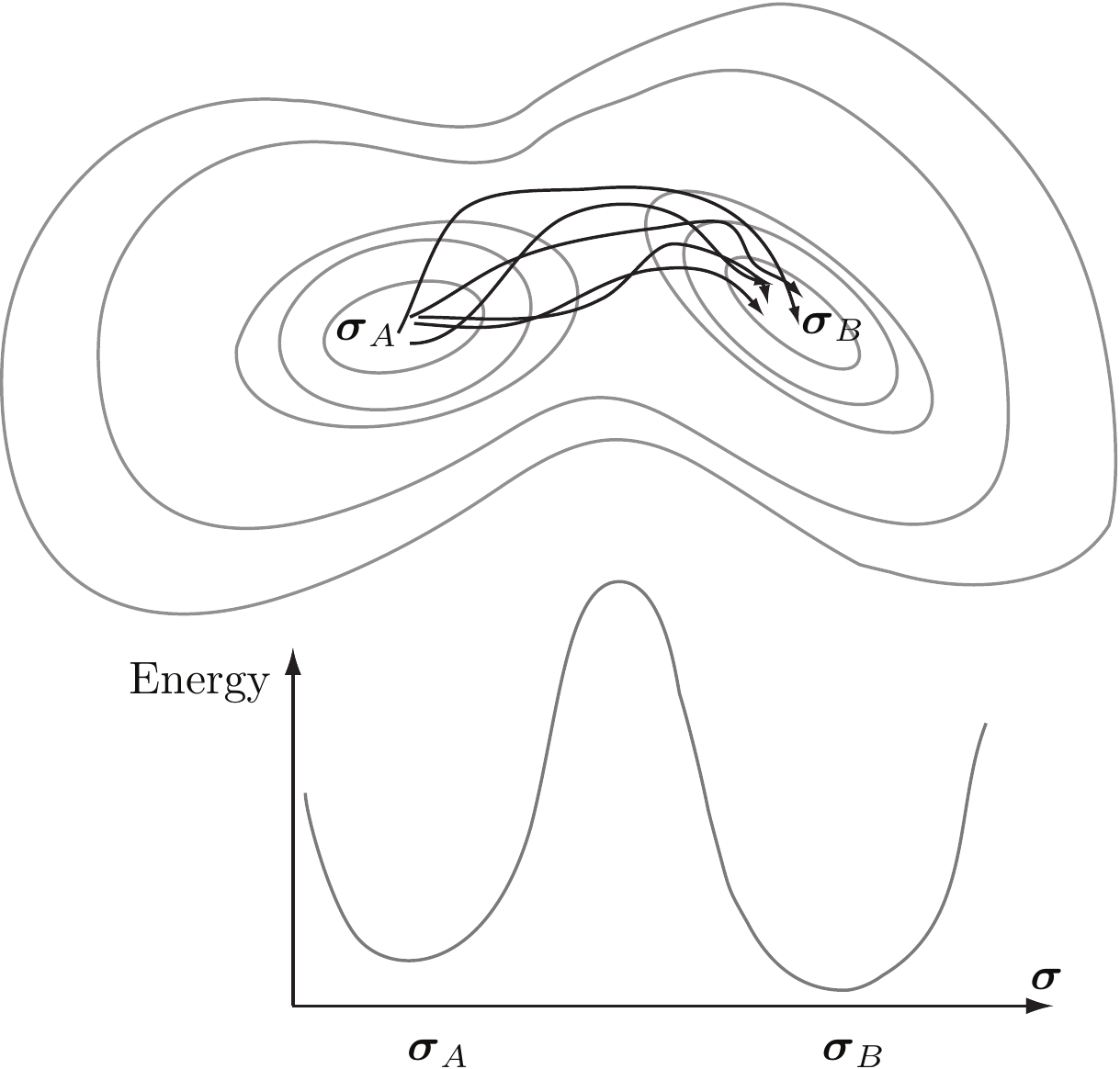}
\caption{A schematic representation of a transition problem between
  two wells. Above: lines of constant energy are represented, as well as
  possible transition paths. Below: side view showing the energy barrier between the two states.}
\label{Figure2}
\end{figure}

Suppose now that the Hamiltonian in Eq.~(\ref{EqHam}) has two deep minima, which we call $A$ and $B$ (see Fig.~\ref{Figure2}).
If we neglect the structure of these minima, at low enough temperature we can write a reduced system with only two states:
\beq
\partial_t \left( 
\begin{array}{c}
p_A(t) \\
p_B(t)
\end{array}
\right) = 
\left( 
\begin{array}{cc}
-k_{A\to B} & k_{B \to A} \\
k_{A \to B} & -k_{B\to A} \\
\end{array}
\right)
\cdot
\left( 
\begin{array}{c}
p_A(t) \\
p_B(t)
\end{array}
\right) 
\eeq
This is of course a gross simplification, but it will prove useful for defining and relating the different quantities that we will consider later.
It is straightforward to check that the evolution operator has one zero eigenvalue (corresponding to the steady-state solution) and one non-zero eigenvalue given by
$\D = k_{A\to B} + k_{B \to A}$, sometimes called ``energy gap" by analogy with quantum mechanics.

The probability to be in $B$ at time $t$ given that the system was in $A$ at time
$t=0$ is given by
\beq\label{ZABtwostates}
Z_{AB}(t) = \frac{k_{A \to B}}{ k_{A\to B} + k_{B \to A} } \big[ 1 - e^{-(k_{A\to B} + k_{B \to A}) t } \big] \ .
\eeq
As we will explain in the following, our method allows us to evaluate $Z_{AB}(t)$ at short times, where $Z_{AB}(t) \approx k_{A\to B} t$, which we will use to extract
the transition rate $k_{A \to B}$. 
It should be noted however that 
once the internal structure of the states $A$ and $B$ is taken into account,
then $Z_{AB}(t)$ is only linear for times larger than
a (small) transient time $\tau_{\rm trans}$: $Z_{AB}(t)\approx k_{A \to B}\times
(t-\tau_{\rm trans})$. This
transient time may be interpreted as the minimal time necessary for
the transition to occur. We will further discuss this point
in the next sections.

As we discussed in the introduction, 
transition rates are usually estimated using a variety of complex 
methods~\cite{valerianirev, chandler, dellago2, tenwoldeffs, weinane_erik, erik2, kurchan1, kurchan2, picciani}. 
We will discuss these at the end of the paper.
For the moment, in order to illustrate the basic difficulty of the problem,
let us discuss three ``naive'' methods that one might try to use to compute $k_{A \to B}$.

The simplest way to estimate transition rates, as well as the full
function $Z_{AB}(t)$, is to recourse to a
traditional Monte-Carlo algorithm, for instance the 
faster-than-the-clock Monte Carlo algorithm
described in details in~\cite[section 7.2.2]{krauth}.
In this case one starts many Monte Carlo simulation in state $A$,
and for each given time $t$ computes $Z_{AB}(t)$ as the fraction
of the simulations that are in state $B$ at time $t$. Clearly, this requires a large
enough number $\NN$ of simulations such that
a sufficient number of trajectories (which is roughly given by $\NN k_{A\to B} t$) perform the jump to state $B$ spontaneously,
a condition which is quite difficult to meet when $k_{A\to B}$ is very small. 
The computational complexity of this method is therefore proportional to $\NN t \propto 1/(k_{A\to B})$, so it scales with the inverse
of the transition rate, which is typically exponential in (some power of) the size of the system.
An example
will be given below, see Fig.~\ref{Fig6}.

Another way is to find the mean first-passage time (MFPT) of transition
from one state to the other, as this time is simply the inverse
transition rate. The MFPT can be calculated
 numerically by solving an equation derived from the backward Master
 equation \cite{gardiner}.
In our simplified two-state model, the probability distribution for the transition time from $A$ to $B$ may be calculated by adding an absorbing boundary condition at $B$. The probability that the system has passed at least once by $B$ after a time $t$, given that it started in $A$, reads:
\beq
Z_{AB}^*(t)=1-e^{-k_{A\to B}t}.
\eeq
The probability distribution function for the time of first passage is then given by $dZ_{AB}^*/dt$, and its mean value is simply the inverse of the transition rate, as expected:
\beq
\text{MFPT}_{A\to B} = \int_{0}^{\infty} dt\, t\,\frac{dZ_{AB}^*}{dt}= \frac{1}{k_{A\to B}}.
\eeq
Note that at short times we have $Z_{AB}(t)\approx Z_{AB}^*(t)\approx k_{A\to B}t$.
Of course, in a generic problem the computational complexity needed 
for the solution of the backward Master equations is proportional to (some power of) the size of the configuration
space of the system, which is typically exponential in the system size (e.g. $2^N$ for a spin system).

The third possibility is to find the energy gap directly by exact diagonalization of the evolution operator ${\cal L}$, by calculating its largest nonzero eigenvalue. 
The gap describes the characteristic rate (inverse of the characteristic timescale) for the equilibration of the system. 
When both states are equiprobable, $k=k_{A\to B}=k_{B\to A}$, the gap is simply $2k$, twice the transition rate. 
Note that when the states are not equiprobable, there is no simple way to infer the transition rates from the gap.
This approach also requires a computational complexity which scales exponentially with the size of the system.

Because each of these ``naive'' methods require a computational time which scales exponentially in the size of the system,
they have a limited span of applicability: Monte Carlo
methods may only sample events that are not too rare; mean first-passage time
and gap calculations
are most efficient for systems with few degrees of freedom.
This is of course the motivation for the development of more sophisticated 
algorithms~\cite{valerianirev, chandler, dellago2, tenwoldeffs, weinane_erik, erik2, kurchan1, kurchan2, picciani}.

\subsection{A simple case: the mean-field model}
\label{sec:MF}

Before proceeding to the description of the numerical method,
it is useful to discuss briefly
the simplest case, namely the mean-field Curie-Weiss
model. This simple, exactly solvable model will help us to set up
notations and get a feeling of the results we should expect for the
two-dimensional system.

The mean-field model corresponds to Eq.~(\ref{EqHam}) with $J_{ij} = 1/(2N)$, and $\tilde h_i = 0$.
It follows from these choices that the Hamiltonian depends only on the global
magnetization $M = \sum_i \s_i$. Therefore, one can reduce the Master equation
acting on the $2^N$ spin configurations to a simpler one that acts only on the $N+1$
possible values of the magnetization $M \in \{ -N, -N+2 \cdots N-2, N
\}$. This allows us
to obtain analytical expressions for the mean first-passage time.

Although these results are not new and have been discussed several times in the literature,
we will discuss them in some details in order to illustrate the problem. Moreover we will
present a  compact derivation that, to our knowledge, has not been previously 
presented in the literature. Here we present the main results, and refer to Appendix \ref{AppendixMF} for details.

We define a free energy at constant magnetization:
\bea
F(M)&=&-\frac{1}{\beta}\log\left[\sum_{\bs|\sum_i\sigma_i=M} e^{-\b H(\bs)}\right]\\
&=&-\frac{1}{\beta}\log\left[\binom{N}{(M+N)/2}e^{\beta M^2/2}\right].
\eea

In the thermodynamic limit, $N\to\io$, we define an intensive free energy:
\beq\label{MFstaticf}
\begin{split}
\b& f(m)  \equiv \lim_{N\to\io}\frac{\b}{N} F(mN)\\
=&- \frac\b2 m^2 +\frac{1+m}2 \log \frac{1+m}2 +  \frac{1-m}2 \log \frac{1-m}2 .
\end{split}
\eeq
Minimization with respect to $m$ gives the thermodynamic free energy.
For $\b < 1$ there is a single minimum at $m=0$. For $\b>1$, there are two minima
at $m=\pm m^*$,
which correspond to two long-lived states at negative and positive magnetization.
We will use those as our states $A$ and $B$, respectively.

Because the Hamiltonian depends only on $M$, 
it follows that at any time $t$,
$p_t(\bs)$ depends only on $M$ as well (provided that this is true at
$t=0$). It is then straightforward to derive a Master equation for
$p_t(M)$ (see Appendix \ref{AppendixMF}).
Transitions rates can then be calculated using standard techniques for
estimating mean first-passage times in one-dimensional systems \cite[Section 7.4]{gardiner}. 

Specifically, one can compute the mean
first passage time in $M_{\rm end}>0$
of a system that starts in $M_{\rm start} < 0$ at time $t=0$. In the
thermodynamic limit, the result does not depend on the start and end
points, as long as they scale linearly with $N$. This mean
first-passage time, which is also the inverse of the transition rate,
reads in this limit:
\beq\label{MF_MFPT_largeN}
\begin{split}
&\text{MFPT}_{A\to B} =\\
&\ \ \frac{\pi}{\b} \sqrt{\frac{1}{[1-\b(1-(m^*)^2)](\b-1)}}e^{\b N [f(0) -
  f(m^*) ]} \ .
\end{split}
\eeq
Besides the prefactor, we recognize Arrhenius law, which relates the
reaction rate to the exponential of height of the free energy barrier.

\begin{figure}
\includegraphics[width =  \linewidth]{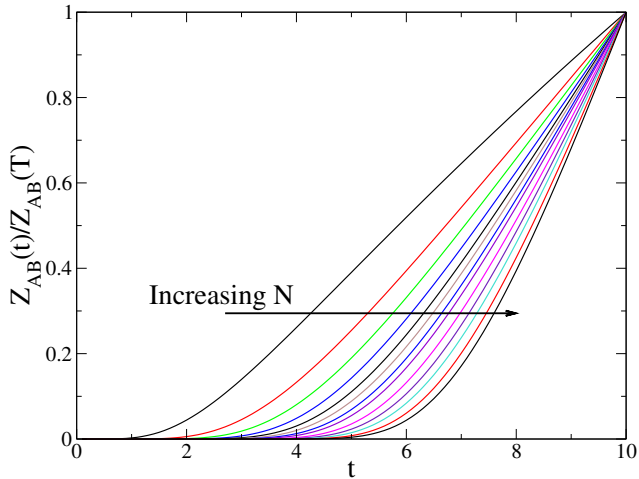}
\caption{
The function $Z_{AB}(t)/Z_{AB}(T)$ (see Appendix \ref{AppendixMF} for details on its calculation), 
for the mean-field model, with $T=10$ and $\b=1.5$, and
for different values of $N=20,40,60,80,100,120,140,160,200,240,280,340,400$ (from left to right).
}
\label{Figure_MF_Zt}
\end{figure}

\begin{figure}
\includegraphics[width =  \linewidth]{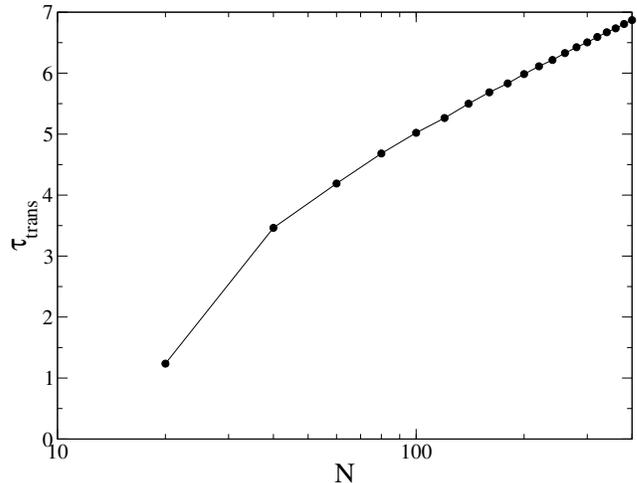}
\caption{
The transient time $\t_{\rm trans}$, as determined by a linear fit $Z_{AB}(t) \sim k_{A\to B} \times (t - \t_{\rm trans})$ of the curves
in Fig.~\ref{Figure_MF_Zt}, is represented as a function of $N$. The
asymptotic behaviour is consistent with the scaling $\sim\log(N)$.
}
\label{Figure_MF_tauN}
\end{figure}

The function $Z_{AB}(t)$ may also be calculated by exact diagonalization of
the evolution operator (see Appendix \ref{AppendixMF}).
The shape of this function at short times is reported in Fig.~\ref{Figure_MF_Zt}.
Keeping only the first two eigenvalues of the evolution operator,
corresponding to the steady state and the gap, one recovers
Eq.~(\ref{ZABtwostates}) in the thermodynamic limit.
However, many other terms are present, which correspond to (much)
larger eigenvalues, and
therefore to much shorter timescales. Due to these terms, the function $Z_{AB}(t)$ is nonlinear at small
$t$; it only becomes linear for times larger than these short time
scales. This nonlinearity is seen in
Figure~\ref{Figure_MF_Zt}. The scaling with $N$ of this time scale is interesting. To determine it,
we fitted $Z_{AB}(t) \sim k_{A\to B} \times (t - \t_{\rm trans})$ at large times (but still much smaller that $1/k_{A\to B}$).
The fit also yields the rate $k_{A\to B}$, which coincides with the one given by Eq.~(\ref{MF_MFPT_largeN}) 
at large $N$. 

The time scale $\t_{\rm trans}$ is related to the time needed to enter the linear regime of $Z_{AB}(t)$.
It is reported in Fig.~\ref{Figure_MF_tauN}, and it scales as $\t_{\rm trans} \propto \log N$ at large $N$.
There is a simple explanation for this. The transition rate is
dominated by the time it takes to climb the barrier up to $M=0$. At the same time, even if the system is prepared
at $M=0$, it takes a time $\sim \log N$ to descend the barrier down to either the positive or negative state.
This can be intuitively justified because, in the $N\to \io$ limit, one can shown that the Master equation is
close to a Fokker-Planck equation with a noise term that scales as $1/N$. It is easy to convice oneself that
in presence of a noise level $\epsilon$, the time it takes to leave an unstable fixed point is of the order of 
$-\log\epsilon$, hence the above scaling follows. We refer the reader to \cite{bakhtin} for a rigorous derivation. 
Therefore, $\log N$ is the minimal time that is needed to cross
the barrier, and it is reasonable to expect $Z_{AB}(t)$ to be
sublinear at these timescales.
This result will turn out to have practical consequences for our
method: in order to 
observe the linear regime of $Z_{AB}(t)$, and extract the transition
rate $k_{A\to B}$, one needs
to be able to compute $Z_{AB}(t)$ for times significantly larger than
the transient time $\tau_{\rm trans}$, which grows with $N$.

\section{Description of the method}\label{sec:method}

Our method relies on the general principles of path sampling, see e.g.~\cite{chandler,francescoprb}.
The idea
is to perform a Monte Carlo sampling of time traces for the
entire system. Each such trajectory is like a configuration in traditional
Monte Carlo methods, and moves in trajectory space are picked randomly
in such a way that the stationary distribution on trajectories coincides with
the desired one \cite{chandler}.
We first lay down the set-up of the problem in the context of the spin
system in section \ref{ss_general}. Then a detailed calculation of the
Monte Carlo transition probabilities are presented in section \ref{ss_weights}.
At this point we are ready to implement the sampling algorithm. We do
this by keeping all spin trajectories fixed, except that of one spin
which is updated as explained in section
\ref{ss_onepath}.
We then describe a procedure for calculating the transition rate from the
sampled trajectories in section~\ref{ss_rates}. 
Following~\cite{chandler}, this method relies on the technique of thermodynamic integration,
as described in section~\ref{AppendixA}.

\begin{figure}
\includegraphics[width =  \linewidth]{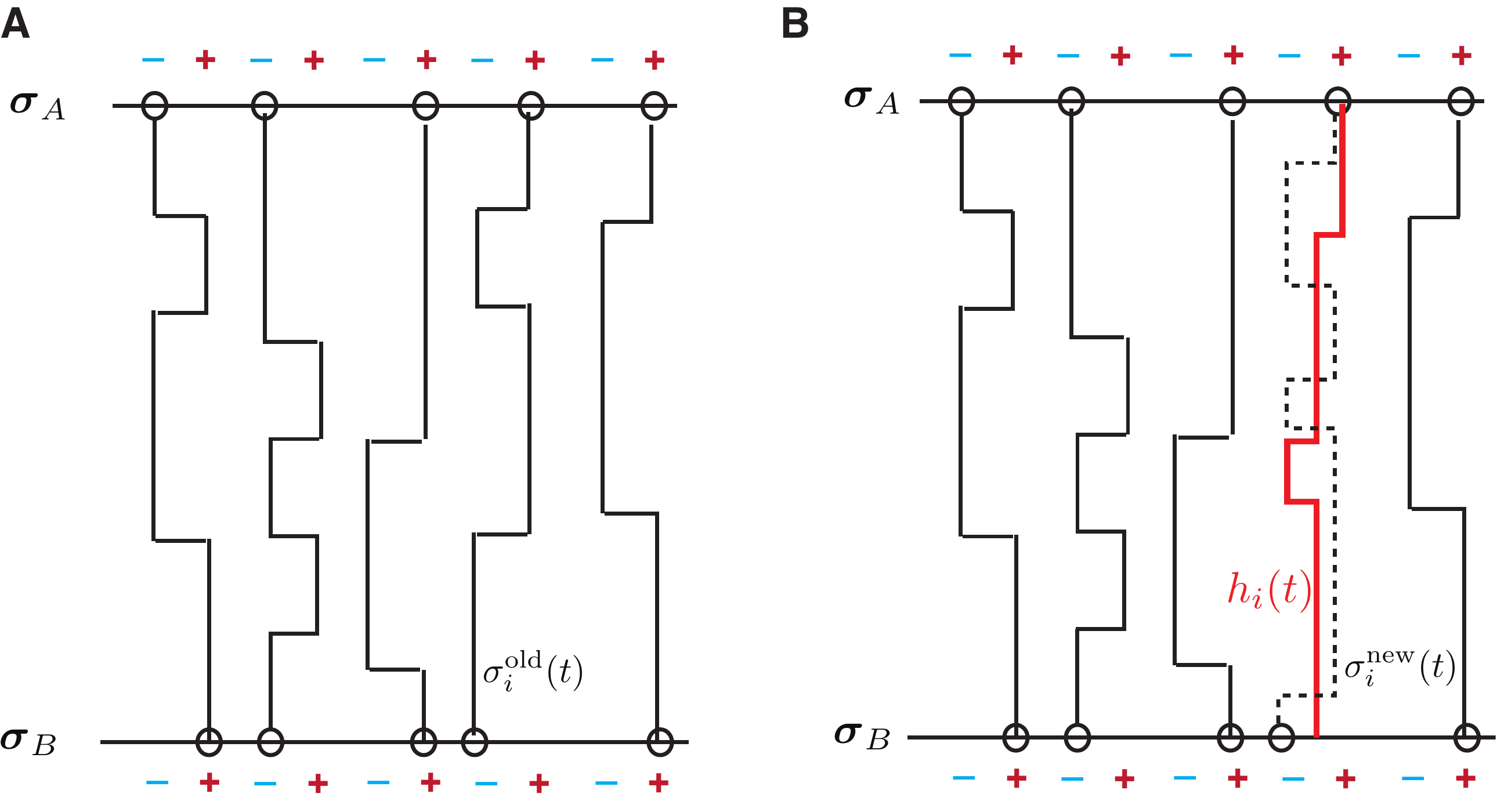}
\caption{{\bf A.} A schematic representation of an $N$-spin trajectory 
  from $\bs_A$ to $\bs_B$. {\bf B.} In the method, we
  choose a spin, erase its trajectory $\sigma_i^{\rm old}(t)$, and
  replace it by an new trajectory  $\sigma_i^{\rm new}(t)$ randomly drawn
  in the effective field $h_i(t)$ created by the other spins, while
  keeping the initial and final conditions fixed.}
\label{Fig3}
\end{figure}

\subsection{General framework}\label{ss_general}

Our goal is to construct an efficient technique for calculating the
escape rates between attractors in a spin system. We
consider all trajectories that start in one attractor $A$ at time $t=0$, and end in
another attractor $B$ at time $t=\TT$---{\em i.e.} all trajectories with fixed
boundaries as depicted in Figure \ref{Figure2}.  We
denote the initial configuration as $\bs_A$ and  the final
configuration as $\bs_B$. To calculate the escape rate, we need to sum
up the normalized probabilities of all possible paths that go between these two points. To do this we will propose a Monte Carlo procedure on trajectories (paths) with fixed boundary conditions (see section \ref{ss_weights}). The results of the sampling can be then integrated numerically to give the transition probabilities between metastable states.

Fig. \ref{Fig3} summarizes the basic idea of our approach to path sampling,
which is analogous to the standard heat bath Monte Carlo algorithm, and was already
applied to a quantum
Monte Carlo algorithm in~\cite{francescoprb}.
We consider a trajectory for
$N$ spins between configuration $A$ and $B$. We want to sample the
space of all possible trajectories. In each Monte-Carlo step, we fix all spins but
one, let us call it $i$. The spins interact with each other via the
Ising Hamiltonian in Eq.~(\ref{EqHam}). If we freeze the trajectories
for all spin but $i$, spin $i$ feels the effect of all the other spins
via an effective time-dependent field $h_i(t)$:
\beq
h_i(t)=\tilde{h}_i+\sum_{j \neq i} J_{i,j} \sigma_j(t).
\eeq

Then, we will redraw (resample) the trajectory for spin $i$, according
to the probability distribution for the spin trajectories with fixed
ends, which is described in section \ref{ss_weights}. We repeat the
procedure by choosing
another spin at random and redraw its trajectory in the same fashion, 
until the system of trajectories
has reached equilibrium.

In section \ref{sec:results} we show that we can sample the space of paths well.
Similarly to Dellago et al.~\cite{chandler}, we use this sampling to compute the
overall normalization of the trajectories which begin in $A$ and end in $B$,  $Z_{AB}(\TT)$,
by means of thermodynamic integration, as described in section~\ref{AppendixA}.
Given this quantity, we can extract the transition rate as discussed above, by means
of a linear fit at large $\TT$.

\subsection{Probability of a path}\label{ss_weights}

We now write the probability for a given path $\bs(t)$ of the system of $N$ spins.
We assign a probability $P_A(\bs_A)$ to the initial state and a weight
$\chi_B(\bs_B)$ on the final state, which will be used to constraint it. Then the probability of a path, in discrete time over
$N_s$ steps (the total time being ${\cal T} = N_s dt$), is
\begin{widetext}
\beq
\mathbb{P}(\bs(t)) = P_A(\bs_A) \prod_{t=dt}^{N_s dt} \left[ 
\left(1 - \sum_{\bs'} w_{\bs';\bs_t} dt \right) \delta_{\bs_t,\bs_{t+dt}}
+ w_{\bs_{t+dt};\bs_t} dt (1-\delta_{\bs_t,\bs_{t+dt}})
\right] \chi_B(\bs_B)
\eeq
\end{widetext}
The first term in the product describes the probability that no spin flips in time $dt$, and the second term accounts for all the possible spin flips that can occur, as described by the rate matrix $w_{\bs_{t+dt};\bs_t}$.

To write the continuum limit of this expression,
we subdivide the trajectory into $m=1,..., M$ intervals,
such that the configuration inside each interval is constant. 
The first interval starts at $t_0 =0$ and $\bs = \bs^1 = \bs_A$ up to $t_1$, the second
interval starts at $t_1$ and ends at $t_2$ and $\bs = \bs^2$, and so on, until the last
interval which starts at $t_{M-1}$ and ends at $t_{M} = {\cal T}$ with $\bs = \bs^{M} = \bs_B$. 
In this case the probability density of a whole trajectory can be written as:
\begin{widetext}
\beq
\text{d}\mathbb{P}(\bs(t))= P_A(\bs_A)
\left\{
\prod_{m=1}^M \exp\left[-(t_m-t_{m-1})\sum_{\bs}w_{\bs;\bs^m}\right] \prod_{m=1}^{M-1} w_{\bs^{m+1},\bs^{m}} dt_m
\right\}
\chi_B(\bs_B)
\ .
\label{probtrajall0}
\eeq
\end{widetext}

The first term describes the probability of nothing happening (no
flip) to any of the spins in a given time interval between $t_m$ and
$t_{m-1}$,
$\exp\left[-(t_m-t_{m-1})\sum_{\bs}w_{\bs;\bs^m}\right]$. The second
term describes the probability of a spin flip happening at the end of
that interval, $w_{\bs^{m+1},\bs^{m}}$. Then we take the product over
all intervals $m=1,..., M$, since the events in each interval are
independent. Note that there are $M$ intervals, but $M-1$ ends of intervals, and that the density $\text{d}\mathbb{P}$ has to be interpreted,
for a given $M$, as a density over the continuous flip times $dt_1 \cdots dt_{M-1}$.

With the choice of the rates we used when writing
Eq.~(\ref{mastereqn}), this expression simplifies greatly, because the
only kind of event
that can happen are single spin flips (only one spin can flip at a time).
We denote
by $i_m$ the spin that flips at time $t_m$.
Therefore $\sigma^{m+1}_{i_m} = -\sigma^{m}_{i_m}$. The rates can be
rewritten as:
\beq
w_{\bs^{m+1};\bs^{m}} =  w(-2 h^{m+1}_{i_m} \sigma^{m+1}_{i_m}) =  w(2 h^{m}_{i_m} \sigma^{m}_{i_m}) \ ,
\label{rate}
\eeq
(note that $h_{i_m}^m=h_{i_m}^{m+1}$, as only $m$ flips between $m$ and $m+1$), and
\beq
\sum_{\bs}w_{\bs;\bs^m} = \sum_i w(2 h^{m}_i \sigma^{m}_i) \ .
\label{rate2}
\eeq
Using Eqs.~(\ref{rate}) and (\ref{rate2}) we can rewrite the probability of the whole trajectory in Eq.~(\ref{probtrajall0}) as:
\begin{widetext}
\beq
\text{d}\mathbb{P}(\bs(t))= P_A(\bs_A)
\left\{
\prod_{m=1}^M \exp\left[-(t_m-t_{m-1})\sum_i w(2 h^{m}_i \sigma^{m}_i)\right] 
\prod_{m=1}^{M-1} w(2 h^{m}_{i_m} \sigma^{m}_{i_m}) dt_m
\right\}
\chi_B(\bs_B)
\ .
\label{probtrajall1}
\eeq
\end{widetext}

\subsection{Updating one spin path}\label{ss_onepath}

\begin{figure}[t]
\includegraphics[width =  \linewidth]{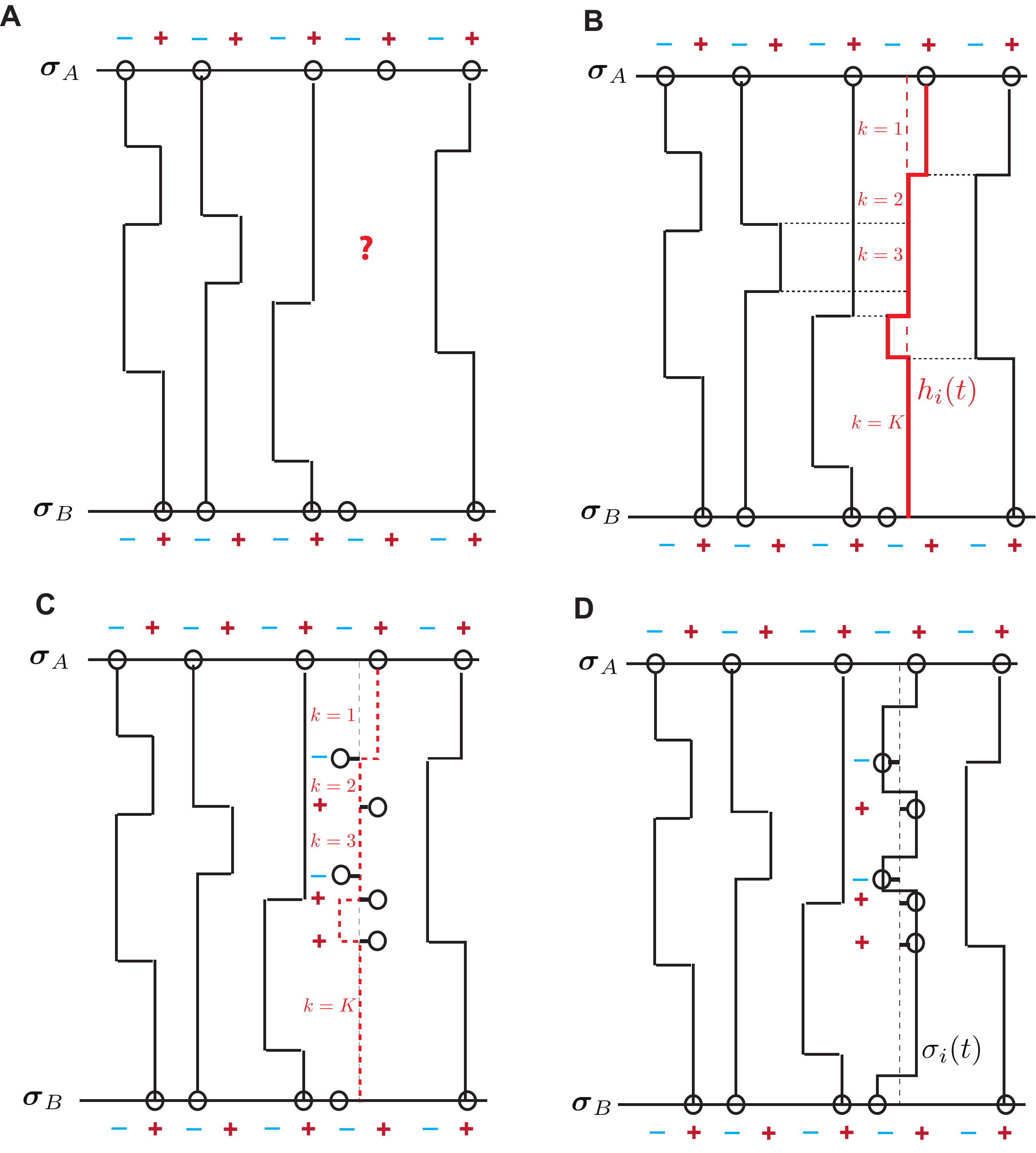}
\caption{Updating one spin path. {\bf A}. We fix all the other spins
  and redraw the trajectory for spin $i$. {\bf B}. The probability of
  a trajectory for spin $i$ depends on the effective external field
  $h_i(t)$ that comes from all the other frozen spins $j$, as well as the
  fields $\bar h_j$ that each of these spins feel in the absence of
  $i$. We divide the trajectory into $K$ time intervals, denoted by
  $k$, on which these fields are constant. {\bf C}. We first draw a
  value of the spins at the boundaries of the $K$ interval, based on
  Eq.~(\ref{eq:probacheckpoints}).  {\bf D}. We then fill in the
  trajectory between these boundaries for each of the intervals,
  according to
  Eqs.~(\ref{insideinterval1})--(\ref{insideinterval2}). 
  }
\label{Fig4}
\end{figure}

 Now, as outlined in section \ref{ss_general}, we fix all spins but
 one, $\sigma_i$, and redraw its trajectory (see Fig.\ref{Fig4} {\bf
   A}). This spin now evolves according to the effective external field
 $h_i(t)$, as shown in Fig.\ref{Fig4} {\bf B}, which varies according
 to the spins with which $i$ interacts. We define $K$ time intervals,
 indexed by $k=1,\ldots,K$, delimited by the times $t_0=0,t_1,\ldots
 t_K=\mathcal{T}$ at which the
 environment of $i$ changes, that is, the times when one of the other spins
 flips (see Fig.\ref{Fig4} {\bf B} ). Let us call $j_k$ the spin that
 flips at time $t_k$. In
 each interval $k$, the spin $i$ sees a constant effective field
 $h^k_i$, as shown in Figure \ref{Fig4}.

\begin{figure}[t]
\includegraphics[width =  \linewidth]{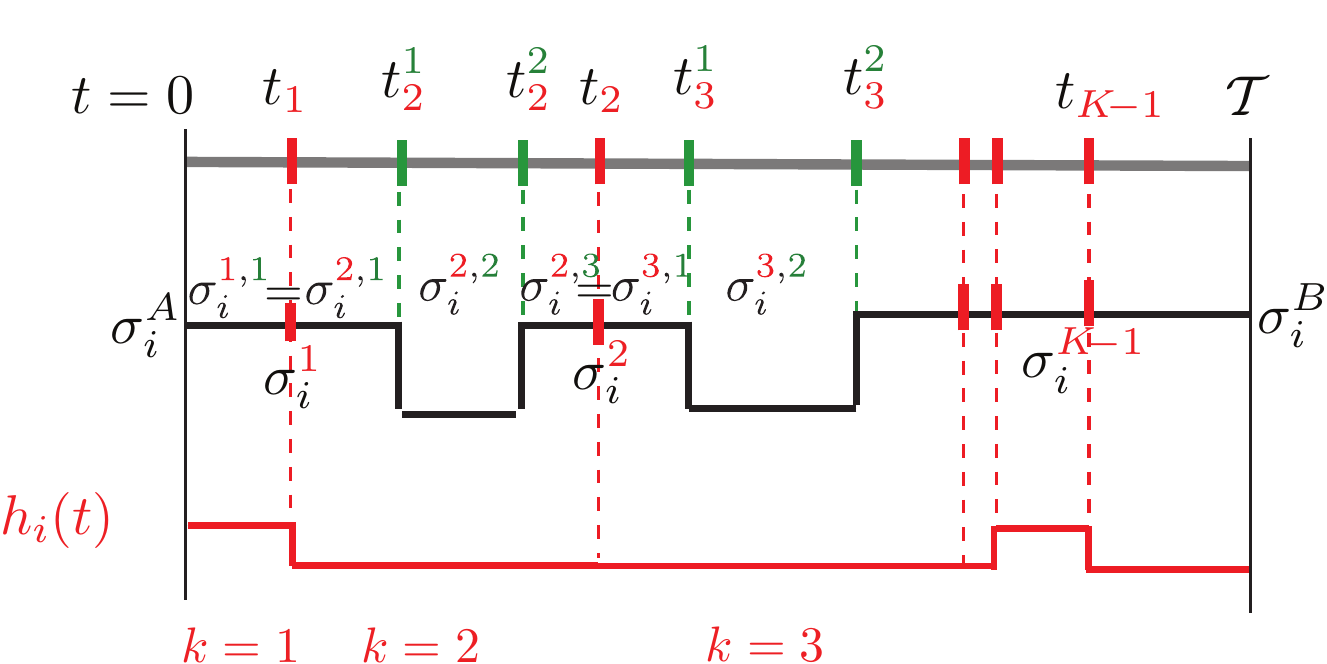}
\caption{Schematic of notations used for drawing a single-spin
  path. The times $t_k$ at which the environment changes are denoted
  by red marks. The value of spin $i$ at these times, $\sigma_i^k$, is
  drawn from Eq.~(\ref{eq:probacheckpoints}). Spin flips of $i$ within each
  interval occur at times $t_k^{\ell}$, denoted by green marks. At
  these times spin $i$ flips from $\sigma_{i}^{k,\ell}$ to $\sigma_{i}^{k,\ell+1}$. 
}
\label{Fig5}
\end{figure}

The conditional
 probability distribution from which the path for spin $i$ is chosen,
 ${\rm d}\mathbb{P}(\sigma_i(t)|\bs_{\backslash i}(t))$, can be
 derived from the expression in Eq.~(\ref{probtrajall1}).
Let us consider each interval $k$ in which the environment of $i$
is constant. 
Within each interval $(t_{k-1},t_k)$, let us define $l_k$
sub-intervals, indexed by $\ell=1,\ldots,l_k$, and delimited by the times
$t^\ell_k$, $\ell=1,\ldots,l_k-1$, defined as the times when spin $i$ flips. We extend
this definition with the convention $t^0_k=t_{k-1}$ and $t^{l_k}_k=t_k$.
The value of spin $i$ in sub-interval $(k,\ell)$ is constant and is
denoted by $\sigma_i^{k,\ell}$. Naturally at $t=t_k$ we have
$\sigma_i^{k,l_k}=\sigma_i^{k+1,1}\equiv \sigma_i^k $.
These notations for intervals, subintervals, and spin values are schematically depicted in Fig. \ref{Fig5}. 

The expression in  Eq.~(\ref{probtrajall1}) has to be broken up into the terms that describe 
the flips of $\sigma_i$, and the terms that describe the evolution of
the other (frozen) spins, 
which depends also on $\sigma_i(t)$, through the effective field that they feel: 
$h_j(t)=\tilde h_j+J_{ji}\sigma_i(t)+\sum_{p\neq \{i,j\}}J_{jp}
\sigma_p(t)$. Isolating the part that depends on $\sigma_i$,  
we can rewrite $h_j(t) = \bar h_j^k + J_{ij} \sigma_i(t)$ where $\bar h_j^k$ is a constant
in each interval $k$.

Putting all this together we can write the conditional probability distributions from which the trajectories for spin $i$ are chosen, keeping the other spins, $j\neq i$ fixed as:
\begin{widetext}
\beq
\begin{split}
\text{d}\mathbb{P}(\sigma_i(t)|\bs_{\backslash i}(t))\propto &
P_A(\sigma_{A,i} | \bs_{A,\backslash i})
\\ \times &
\prod_{k=1}^K 
\Big\{ \prod_{\ell=1}^{l_k} \exp\big[ -(t^{\ell}_k-t^{\ell-1}_{k})
[ w(2 h_i^k \sigma_i^{k,\ell}) + \sum_{j\neq i} w(2  \sigma_j^{k}(\bar h_{j}^{k} + J_{ji} \sigma^{k,\ell}_i))]
\big]
\prod_{\ell=1}^{l_k-1}w(2h_i^k \sigma_i^{k,\ell}) dt_{k,\ell}
\Big\}
\\  \times &
\prod_{k=1}^{K-1}w[ 2 \sigma_{j_k}^{k} (\bar h_{j_k}^{k} + J_{j_ki} \sigma^{k}_i)  ] \
\chi_B(\sigma_{B,i} | \bs_{B,\backslash i}) \ 
.
\label{condprob1}
\end{split}
\eeq
\end{widetext}

The term in the curly brackets describes the evolution of $\sigma_i$ in one of the intervals of constant environment,
which has two contributions:
\begin{enumerate}
\item
 The product of exponentials comes from inside the sub-intervals, where neither
 $\s_i$ or its environment change. It has itself two contributions: one is
  the probability of $i$ not flipping, the other is the probability of
  all other spins not flipping.
\item
 The second product in the curly brackets is the probability of
 $\sigma_i$ flipping, which happens between each $(k,\ell)$ subinterval.
\end{enumerate}
The third line and last product over $k$ is the probability of 
spin $j_k$ flipping at time $t_k$, which depends on $\sigma_i$ through
the field $h_{j_k}$. Note that the rate of flipping depends on the value of $\sigma_i^{k,l_k} = \sigma_i^k$ 
specifically at the end of the $k$ interval.

We now want to draw a single-spin trajectory $\sigma_i(t)$ from the
probability distribution described by Eq.~(\ref{condprob1}). 
Following~\cite{francescoprb}, we split
this task in two parts.
First we draw the values $\sigma_i^k$ of spin $i$ at the boundary times $t_k$
(section \ref{ss_onepath_1}). Second we draw the trajectory of $\sigma_i(t)$ in each of
the intervals $(t_{k-1},t_k)$ with fixed initial and final conditions
$\sigma_i^{k-1}$ and $\sigma_i^k$, which just amounts to drawing the
times $t^k_\ell$ (section \ref{ss_onepath_2}).

\subsubsection{Drawing the boundary values}\label{ss_onepath_1}

Here we show how one can draw the values of spin $i$ between intervals of constant environment, 
denoted by $\sigma_i^k = \sigma_i^{k,l_k}=\sigma_i^{k+1,1}$, 
together with the initial and final values $\sigma^0_i\equiv
\sigma^A_{i}$ and $\sigma^{K}_i\equiv \sigma^B_{i}$.
Having fixed the values at the boundaries of the $k$ intervals, 
we will then draw the trajectory for $\sigma_i$ in each interval $k$. 

We therefore have to construct the joint probability 
$\mathbb{P}(\sigma^A_{i}=\sigma^0_i, \sigma_i^{1}, \sigma_i^{2}, \cdots, \sigma_i^{K}=\sigma^B_{i})$
of the boundary values of $\sigma_i$.
To do this we need to sum over all possible paths that are consistent with the given boundary values.
This is easily done by considering the terms in curly brackets in Eq.~(\ref{condprob1}), and observing that
its sum over paths going from $\sigma_i^{k-1}$ to $\sigma_i^k$ can be written as follows:
\begin{widetext}
\beq
\sum_{\text{paths}}^{\sigma_i^{k-1} \to \sigma_i^k} \left\{ \prod_{\ell=1}^{l_k} \exp\big[ -(t^{\ell}_k-t^{\ell-1}_{k})
[ w(2 h_i^k \sigma_i^{k,\ell}) + \sum_{j\neq i} w(2  \sigma_j^{k}(\bar h_{j}^{k} + J_{ji} \sigma^{k,\ell}_i))]
\big]
\prod_{\ell=1}^{l_k}w(2h_i^k \sigma_i^{k,\ell}) dt_{k,\ell} \right\}= 
\bra{\sigma_i^k} e^{(t_k-t_{k-1}) {\cal L}_i^k)}
\ket{\sigma_i^{k-1}}
\eeq
where the operator ${\cal L}^k_i$ is a $2\times2$ matrix defined by
\beq
\bra{\sigma'} {\cal L}^k_i \ket{\sigma} = 
\left\{
\begin{array}{ll}
   -w(2 h_i^k \sigma) - \sum_{j\neq i} w(2  \sigma_j^{k}(\bar h_{j}^{k} + J_{ji} \sigma ))   
 & \text{ for } \sigma' = \sigma \\
   w(2h_i^k \sigma)                 & \text{            for } \sigma' = -\sigma
\end{array}
\right.
\eeq
\end{widetext}
This relation is formally equivalent to a ``Suzuki-Trotter''
representation \cite{suzukitrotter}, and may be obtained by
discretizing in small time steps $dt$ and expanding the
exponentials. For a detailed derivation of a similar relation, see \cite{francescoprb}.
Note that the matrix ${\cal L}_i^k$ differs from the transition rate matrix for a spin evolving in a constant field:
indeed, we note that having fixed (frozen) all the other spins $j$, we interfered with the natural dynamics of 
the system, and we cannot now derive the probability of the trajectory for spin $i$ directly 
from collapsing the Master equation. 
Still we can interpret the result above as if the spin $i$ was evolving under the modified Markov dynamics
\beq
\partial_t p_t={\cal L}_i^k p_t \ .
\eeq
However, this analogy might be misleading since $\sum_{\sigma'}\bra{\sigma'} {\cal L}^k_i \ket{\sigma}  \neq 0$,
therefore the dynamics does not conserve the probability (the vector
$p_t$ cannot be interpreted as a probability).

In order to find the density distribution from which the values of $\sigma_i^k$ are drawn 
(the values at the boundaries of the $k$ intervals), we use this result and 
we obtain:
\beq\label{eq:probacheckpoints}
\begin{split}
\mathbb{P}(\{\sigma_i^k\} | \bs_{\backslash i}(t))
\propto & e^{h^A_{i} \sigma^A_{i} } 
\prod_{k=1}^K \bra{\sigma_i^k} e^{(t_k-t_{k-1}) {\cal L}_i^k)}
\ket{\sigma_i^{k-1}}
\\
\times &
\left\{\prod_{k=1}^{K-1}w[ 2 \sigma_{j_k}^{k} (\bar h_{j_k}^{k} + J_{j_k i}
\sigma^{k}_i)  ]\right\}  e^{h^B_{i} \sigma^B_{i} } \ ,
\end{split}
\eeq
The weight on the boundary states $A$ and $B$ are described by effective
fields $h^{A}_i$, $h^B_i$, which depends on the other spins (this is
possible because spins can take only two values).
The first product is the probability of transitioning from
$\sigma_i^{k-1}$ to $\sigma_i^{k}$ in interval $(t_{k-1},t_k)$, 
and the second product contains the dependencies of the other spin
flips on $\sigma_i$. 
In the form written above in Eq.~(\ref{eq:probacheckpoints}), $\mathbb{P}(\{\sigma_i^k\} | \bs_{\backslash i}(t))$
is a one-dimensional Ising chain, therefore the values of $\{\sigma_i^k\}$ can be easily drawn by means of transfer
matrices \cite{francescoprb}.

Now one needs to diagonalize the matrix  ${\cal L}$. In our specific example, we can rewrite the matrix ${\cal L}$ in a more compact form:
\beq
\begin{split}
& \bra{\sigma'} {\cal L}^k_i \ket{\sigma} = e^{\b h_i^k \sigma'/2} \bra{\sigma'} {\cal M}^k_i \ket{\sigma} e^{-\b h_i^k \sigma/2} \\
& \bra{\sigma'} {\cal M}^k_i \ket{\sigma} =
\left\{
\begin{array}{ll}
   -w(2 h_i^k \sigma) - \sum_{j\neq i} w(2  \sigma_j^{k}(\bar h_{j}^{k} + J_{ji} \sigma ))   
 \\ \hspace{.5\linewidth} \text{ for } \sigma' = \sigma \\
   w(2h_i^k \sigma)      e^{\b h_i^k \sigma}           \qquad \text{for } \sigma' = -\sigma
\end{array}
\right.
\end{split}
\eeq
Thanks to the detailed balance condition the diagonalization task is simplified. The matrix ${\cal M}$ is
symmetric and can be written as~\footnote{
For non-equilibrium systems, the matrix $M^k_i$ is not symmetric,
but this does not prevent one to use this method.
}
\beq
{\cal M}^k_i = M_i^k \, I + B_i^k \sigma_z + \Gamma_i^k \sigma_x \ ,
\eeq
where
\beq
\Gamma_i^k =w(2h_i^k )      e^{\b h_i^k }  \ ,
\eeq
and
\beq
\begin{split}
 M_i^k =  & \frac12 \left[  -w(2 h_i^k) - \sum_{j\neq i} w(2
   \sigma_j^{k}(\bar h_{j}^{k} + J_{ji} ))\right. \\
&\left. -w(-2 h_i^k) - \sum_{j\neq i} w(2  \sigma_j^{k}(\bar h_{j}^{k}
  - J_{ji} )) \right] \ ,
\end{split}
\eeq
and
\beq
\begin{split}
 B_i^k = &  \frac12 \left[  -w(2 h_i^k) - \sum_{j\neq i} w(2
   \sigma_j^{k}(\bar h_{j}^{k} + J_{ji} )) \right.
\\ & \left.+w(-2 h_i^k) +\sum_{j\neq i} w(2  \sigma_j^{k}(\bar h_{j}^{k} - J_{ji} )) \right] \ ,
\end{split}
\eeq
where $I$ is the identity and $\sigma_{x}, \sigma_{z}$ are Pauli matrices.
The diagonalization of $B \s^z + \Gamma \s^x$ leads to
\beq
\begin{split}
& \langle \s' | e^{\lambda (M \, I + B \s^z + \Gamma \s^x ) }| \s
\rangle = e^{\lambda M}\\
&\qquad
 \times
\begin{cases} 
  \cosh(\lambda\Delta) + \s \frac{B}{\Delta} 
\sinh(\lambda\Delta)  & \text{if} \ \s=\s' \\
\frac{\Gamma}{\Delta} \sinh(\lambda\Delta) & \text{if} \ \s=-\s'
\end{cases} \ ,
\end{split}
\label{eq_propagator_res}
\eeq
with the short-hand $\Delta=\sqrt{B^2+\Gamma^2}$.
We arrive at the final result
\beq
\begin{split}
 \bra{\sigma'} e^{\lambda  {\cal L}^k_i} \ket{\sigma} =& e^{\b h_i^k \sigma'/2} \bra{\sigma'} e^{\lambda {\cal M}^k_i} \ket{\sigma} e^{-\b h_i^k \sigma/2} \\ 
 =&  e^{\b h_i^k (\sigma' -\sigma)/2}  \langle \s' | e^{\lambda (M_i^k \, I + B_i^k \s^z + \Gamma_i^k \s^x ) }| \s \rangle  \\ 
 =&  e^{\lambda M_i^k} \cosh(\lambda\Delta_i^k)  \\
&\times\begin{cases} 
 1 + \s \frac{B_i^k}{\Delta_i^k} 
\tanh(\lambda\Delta_i^k)  &\text{if} \ \s=\s' \\ 
 e^{-\b h_i^k \sigma}\frac{\Gamma_i^k}{\Delta_i^k} \tanh(\lambda\Delta_i^k) &\text{if} \ \s=-\s'
\end{cases} \ .
\end{split}
\eeq
Using this expression, the boundary values $\sigma_i^k$ are drawn according to Eq.~(\ref{eq:probacheckpoints})
using the transfer matrix technique. Note that the constant term
$ e^{\lambda M} \cosh(\lambda\Delta)$ does not depend on $\sigma$ and
can be absorbed into the normalization of Eq.~(\ref{eq:probacheckpoints}).

\subsubsection{Drawing the trajectory inside each interval}\label{ss_onepath_2}

During each interval $(t_{k-1},t_k)$ the values of $M, B, \Gamma,
\Delta, h$ are constant and we drop the indices from now on.
The trajectory of $\sigma_i$ is built recursively. Suppose that the
trajectory has been built up to $t$, $t_{k-1}\leq t<t_k$, and ends at
$\sigma_i(t)= \sigma$ (at the start of the algorithm $t=t_{k-1}$).
We denote by $\lambda=t_k-t$ the duration of the remaining interval, and
$\sigma'\equiv \sigma^k$.
If $\s=\s'$, the probability of $\s_i$ not flipping at all in the remaining interval
$(t,t_k)$ is
\beq\label{insideinterval1}
\frac{e^{\lambda \bra{\sigma}  {\cal L} \ket{\sigma} }}{  \bra{\sigma} e^{\lambda  {\cal L}} \ket{\sigma}  }= \frac{e^{\lambda B \sigma}}{ \cosh(\lambda\Delta) + \s \frac{B}{\Delta} 
\sinh(\lambda\Delta) }.
\eeq
If this is the case, the whole trajectory between $t_{k-1}$ and $t_k$
is now completed and the routine is stopped.
Otherwise, the next flipping event occurs at time $t+u$, where $u$ is
drawn from the cumulative distribution:
\beq\begin{split}
G(u; \sigma, \sigma') =& \frac{ \int_0^u dv \, e^{v  \bra{\sigma}  {\cal L} \ket{\sigma}} \bra{\sigma'} e^{(\lambda-v)  {\cal L}} \ket{-\sigma} w(2 h \sigma) }
{  \int_0^\lambda dv \, e^{v  \bra{\sigma}  {\cal L} \ket{\sigma}} \bra{\sigma'} e^{(\lambda-v)  {\cal L}} \ket{-\sigma}  w(2 h \sigma) }\\
=  &\frac{ \int_0^u dv \, e^{v  B \sigma } \bra{\sigma'} e^{(\lambda-v) ( B \sigma_z + \Gamma \sigma_x )   } \ket{-\sigma} }
 { \int_0^\lambda dv \, e^{v  B \sigma } \bra{\sigma'} e^{(\lambda-v)
     ( B \sigma_z + \Gamma \sigma_x )   } \ket{-\sigma} }.
\end{split}
\eeq

This formula coincides exactly with that in
\cite{francescoprb}, and a short calculation gives:
\beq\label{insideinterval2}\begin{split}
&G(u;\sigma,-\sigma) = 1 - e^{\sigma B u} \frac{\sinh( (\lambda-u) \Delta)}{\sinh( \lambda \Delta)} \\
 &G(u;\sigma,\sigma) = \frac{  
 \cosh( \lambda \Delta) + \frac{\sigma B}{\Delta} \sinh( \lambda \Delta)
}{  
 \cosh( \lambda \Delta) + \frac{\sigma B}{\Delta} \sinh( \lambda \Delta)
-e^{\sigma B \lambda}  }
\\ 
&\quad  -\frac{e^{\sigma B u} [ \cosh( (\lambda-u) \Delta) + \frac{\sigma B}{\Delta} \sinh( (\lambda-u) \Delta) ]}{  
 \cosh( \lambda \Delta) + \frac{\sigma B}{\Delta} \sinh( \lambda \Delta)
-e^{\sigma B \lambda} }.
\end{split}\eeq
Once $u$ is drawn, we update $t\to t+u$, $\sigma\to -\sigma$, and we
repeat the procedure until the trajectory is completed over
$(t_{k-1},t_k)$. We implement this algorithm for each interval.


\subsection{The calculation of the rates}\label{ss_rates}

We assume from now on that, thanks to the algorithm previously described, we are able to sample
efficiently the dynamical trajectories for the whole system generated by the probability in Eq.~(\ref{probtrajall1}),
which we write in a compact form as
\beq\label{PAB}
\text{d}\mathbb{P}(\bs(t))= P_A[\bs(0)] \, \PP[\bs(t)] \, \chi_B[\bs(\TT)] \ .
\eeq
The first term is the probability of the initial condition, the second term describes the stochastic evolution of the system,
the last term is a constraint on the final state. We also indicated explicitly the time $\TT$ 
at which the constraint on $B$ is imposed.
We define the ``partition function'':
\beq
Z_{AB}(\TT) = \int \text{d}\mathbb{P}[\bs(t)] = \sum_{\bs(t)} P_A[\bs(0)] \, \PP[\bs(t)] \, \chi_B[\bs(\TT)] \ ,
\eeq
which is the probability that the system, starting in $A$ at time $t=0$, is found in state $B$ at $t=\TT$; this
was introduced and computed for the reduced two-state problem in Eq.~(\ref{ZABtwostates}) above. This is the quantity
we want to compute in order to extract the transition rate $k_{A\to B}$.
Note that in absence of the constraint at the final time, $\chi_B(\bs) =1$, we have
\beq
Z_{A}(\TT) = \sum_{\bs(t)} P_A[\bs(0)] \, \PP[\bs(t)] = 1 \ ,
\eeq
as follows from the normalization of probability.

\subsubsection{Thermodynamic integration}\label{AppendixA}

The estimation of the $Z_{AB}(\TT)$
requires to use the technique of thermodynamic integration. 
In this technique one chooses a suitable parameter $\mu$ of the system
(e.g. the temperature or the magnetic field: we will give an example below)
and defines an interpolation path $\mu(s)$, $s\in [0,1]$, such
that for $s=0$, $Z_{AB}(\TT, \mu(0))$ can be easily computed, and
that for $s=1$,  $Z_{AB}(\TT, \mu(1))$ coincides with the actual partition function
one wants to estimate. 
Then one carries out the path sampling procedure described in the
previous sections along the interpolation path.
The partition function is then estimated by
\beq
Z_{AB}(\TT,\mu(1)) = Z_{AB}(\TT,\mu(0)) \, \, e^{\int_0^1 ds \, U_{AB}(\TT,\mu(s)) \frac{d\mu}{ds}} \ ,
\eeq
where
$Z_{AB}(\TT,\mu(0))$ is assumed to be 
easily calculable, and where
\beq
U_{AB}(\TT,\mu) = \frac{\partial \log Z_{AB}(\TT,\mu)}{\partial \mu},
\eeq
can be estimated as an average over the transition paths generated at the value $\mu$ of the parameter.

As in usual Monte Carlo methods, the choice of the optimal interpolation path depends on the system
under investigation. Different choices can lead to very different performances of the method, in particular 
because one must avoid the presence of phase transitions along the interpolation path. We will discuss
this problem in more details on our specific example in the following. 

This method may not seem very efficient because one is required to perform a thermodynamic
integration for each value of the final time $\TT$. A large enough number of values of $Z_{AB}(\TT)$ are indeed required
to identify the large-time linear regime and extract $k_{A\to B}$, as
we have already discussed. Luckily enough, in some cases
one can avoid performing these multiple thermodynamic integrations thanks to a trick introduced by
Dellago et al.~\cite{chandler}, which we discuss in the next section.

\subsubsection{An approximated method to compute the time dependence of $Z_{AB}(\TT)$}
\label{sec:fullZt}

Following Dellago et al.  \cite{chandler}, we notice that if $\TT$ is much shorter than the transition time $1/k_{A\to B}$, and if $\t < \TT$ we can write
\beq\begin{split}
&Z_{AB}(\t)  = \sum_{\bs(t)} P_A[\bs(0)] \, \PP[\bs(t)] \, \chi_B[\bs(\t)]  \\
&\ \approx \sum_{\bs(t)} P_A[\bs(0)] \, \PP[\bs(t)] \, \chi_B[\bs(\t)] \, \chi_B[\bs(\TT)] \\
&\  = Z_{AB}(\TT) \, \frac{\sum_{\bs(t)} P_A[\bs(0)] \, \PP[\bs(t)] \, \chi_B[\bs(\t)] \, \chi_B[\bs(\TT)]}{\sum_{\bs(t)} P_A[\bs(0)] \, \PP[\bs(t)] \, \chi_B[\bs(\TT)]} \\
&\  = Z_{AB}(\TT) \, \la  \chi_B[\bs(\t)]  \ra_{AB,\TT} \ ,\label{smallertime}
\end{split}\eeq
where $\la\bullet\ra_{AB,\TT}$ denotes an average over the path probability
measure in Eq.~(\ref{PAB}). The approximation made here is that the system
does not transition
back to state $A$ at time $\TT$ if it has reached state $B$ at
$\tau<\TT$ (in other words, the system may transition only once in a short enough time).

Because we are able to sample efficiently from this probability measure,
computing $ \la  \chi_B[\bs(\t)]  \ra_{AB,\TT} $  initially is expected to be 
an easy task (but we will see in the following that this is not always the case). 
Indeed, $ \la  \chi_B[\bs(\t)]  \ra_{AB,\TT} $  is the probability that
the system is in state $B$ at time $\tau$ given that it was in state
$A$ initially and that it will reach state $B$ at time $\TT$. This
probability can be estimated by examining our sampled paths from $A$
to $B$ and ask what fraction has already reached $B$ at times
$\tau<\TT$.

In this way, one can perform a single thermodynamic integration to measure 
$Z_{AB}(\TT)$ for a large enough time $\TT$, and then use the trick described above
to obtain $Z_{AB}(\t)$ for all $\t \leq \TT$ from a single path Monte Carlo simulation at
the target value of the parameters.

\section{Application to the 2D ferromagnetic Ising model}\label{sec:results}

In this section we apply the path-sampling Monte-Carlo algorithm described above to a
specific example---the two-dimensional ferromagnetic Ising model. 
We start by presenting a few technical checkpoints that ensure that
our sampling algorithm is working well, 
 and we then present the results for
the transition rate.
We then discuss them in the light of known results on the surface
tension and theoretical arguments~\cite{gallavotti,martinelli}.
Note that nucleation problems in this model have been already studied
by a number of methods~\cite{venturoli,shearising}.

The 2D Ising model is defined by Eq.~(\ref{EqHam})
with $J_{ij}=J=1$ (without loss of generality) for neighboring spins on a 
square lattice 
containing $N=L^2$ sites 
with periodic boundary conditions, and $J_{ij}=0$ otherwise. 
Note that an important simplification of our method is made possible by the sparseness
of interactions between spins. In general the
intervals $k=1,\ldots,K$ are delimited by events where any other spin than $i$
is flipped. Here we can restrict this definition to nearest and
second-nearest neighbors spins, because neither $h_i(t)$ nor $\bar h^k_j$
are affected by more distant spins being flipped.

In the absence of an external field, $\tilde{h}_i =0$, 
this model has two deep energy minima where all spins are up or down,
and below some critical temperature $T_c=1/\b_c$, {\em i.e.} for
$\beta>\beta_c=\log(1+\sqrt{2})/2\approx 0.4407$, the system 
 at equilibrium is typically found close to
one of these two minima, called $A$ and $B$ \cite{onsager,gallavotti}. 
The free energy barrier separating the two minima is expected to be of the order
of $L$~\cite{onsager,gallavotti,martinelli}, as we will discuss in more details below.
Due to the
symmetry of the model, $k_{A\to B}=k_{B\to A}$ and the energy gap is equal to
twice the escape rate from state $A$ to state $B$, 
which is expected to be of
the order of $\exp(-L)$.
We impose
the initial and final states by setting:
\bea
P_A(\bs_A)&=&\frac{\exp\left(h_A M_A\right)}{(2\cosh h_A)^N} 
= \prod_{i=1}^N \frac{e^{h_A \s_i^A}}{2\cosh h_A}
\\
\chi_B(\bs_B)&=&\exp\left[-h_B \left(M^*-M_B \right)
\theta \left(M^*-M_B \right)\right]\nonumber
\eea
with as usual $M=\sum_{i=1}^N\sigma_i$, and $\theta(x)$ is the Heaviside function.

All the simulations we report below have been performed at a temperature $\b=1$, which
is well below the critical temperature and correspond to an equilibrium magnetization
per spin
$m_{\rm eq} = 0.999275\ldots$ according to the Onsager formula~\cite{onsager}. Therefore,
the two states $A$ and $B$ are very concentrated around the configurations with all
spin up or all spin down.

We have chosen $h_A=-3$ and $h_B=1$, $M^* = [0.56 N]$ so that the system starts in the down
state (which we call $A$ from now on) 
and finishes in the up ($B$) state. We have checked that the precise values of these parameters
are irrelevant for the determination of the transition rate.

\subsection{Numerical results}

\subsubsection{Thermodynamic integration}

As previously discussed, to compute $Z_{AB}(\TT)$ we must use thermodynamic integration
over a parameter $\mu$.
We have at least two possibilities:
\begin{itemize}
\item We choose $\mu = h_B$. We start at $h_B=0$, where the system has no constraint on
the final state: there $Z_{AB}(\TT)=1$. Then we change $h_B$ from $0$ to the final value
$h_B = 1$.
\item We choose $\mu = \b$. We start at $\b=0$, i.e. 
at infinite temperature where the dynamics of the spin is decoupled. Then we change the
temperature from $\b=0$ to the final temperature $\b=1$.
\end{itemize}
Although for small sizes we can use both strategies (and checked that
we get fully compatible results),
the first strategy is not efficient at large sizes. The reason is that at $h_B=0$ the 
system has no constraint on the final state, and therefore for small $\TT$ it will be
typically in state $A$. On the contrary, at $h_B=1$, the constraint is strong and the
system will be typically in state $B$. We found that the system of paths undergoes a first
order phase transition as a function of $h_B$ along the integration path from $h_B=0$ to
$h_B=1$, which is somehow similar to the first order transition that the standard spin system
undergoes as a function of the external field below $T_c$. Around this transition, hysteresis
is observed and equilibrating the path system becomes extremely difficult, thus spoiling the
efficiency of the algorithm. Therefore, in the following, we abandon the first strategy and
only focus on the second one, for which this problem is absent~\footnote{We thank 
P.~Charbonneau for a crucial discussion on this point.}.

Before discussing the second strategy
we wish to stress that the
first strategy is the one that was used in the original paper of Dellago et al.~\cite{chandler};
and it worked only because the investigated system was extremely small. In general, we speculate
that doing the thermodynamic integration on the constraint on the final state $\chi_B$ will 
always produce this problem for large enough systems.

We therefore now discuss in more details the thermodynamic integration in temperature.
At infinite temperature, $\b=0$, the spins are decoupled and undergo independent Glauber dynamics
in absence of any external field. Therefore, it is easy to show that for a single spin,
\beq
p_{\rm up}(t) = \frac12 \left[ 1 + (2 p_{\rm up}(0) - 1) e^{-2 t} \right]
\eeq
with $p_{\rm up}(0) = e^{h_A}/(2 \cosh(h_A))$. The probability that the system has magnetization $M$ at time
$t$ is:
\beq
P_t(M) = \binom{N}{\frac{N+M}2} p_{\rm up}(t)^{(N+M)/2} (1-p_{\rm up}(t) )^{(N-M)/2} \ ,
\eeq
and the partition function at $\beta=0$ is therefore:
\beq
Z_{AB}(\TT,\beta=0) = \sum_M P_\TT(M) e^{-h_B (M^* - M) \theta(M^* - M)} \ ,
\eeq
that can be numerically computed very easily for any $N$.

\begin{figure}
\includegraphics[width =  \linewidth]{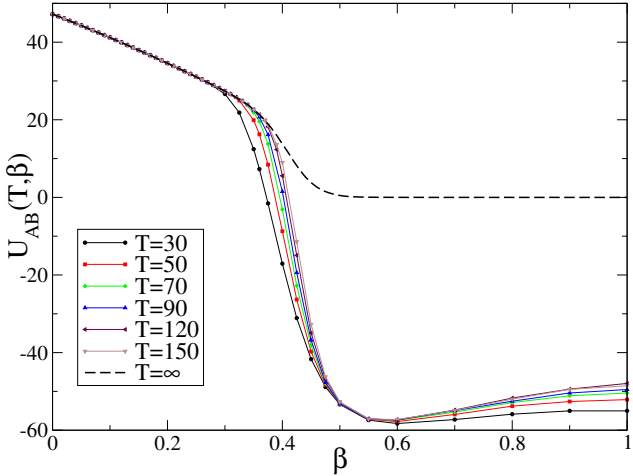}
\caption{An example of the thermodynamic integration procedure. The quantity $U_{AB}(\TT,\b)$ 
defined in Eq.~(\ref{thermint})
is reported as a function of the inverse temperature $\beta$ for several values of $\TT$, 
for $N=81$ (here $h_{A}=-3$, $h_{B}=1$, $M^*=45$).
}
\label{Fig8}
\end{figure}

Next, we need the derivative of $Z_{AB}(\TT,\b)$ with respect to $\b$. 
A straightforward calculation starting from Eq.~(\ref{probtrajall1})
gives:
\begin{widetext}
\beq\begin{split}
&U_{AB}(\TT,\b) = \frac{\partial \log Z_{AB}(\TT,\b)}{\partial \b} \\
& = \frac{1}{Z_{AB}(\TT,\b)} \int \text{d}\mathbb{P}(\bs(t))
\left\{
\sum_{k=1}^L (t_k-t_{k-1}) \sum_i w(2 h^{k}_i \sigma^{k}_i) h^{k}_i \sigma^{k}_i- \sum_{k=1}^{L-1} h^{k}_{i_k} \sigma^{k}_{i_k}
\right\} \\
& = \left\la \sum_{k=1}^L (t_k-t_{k-1})\sum_i w(2 h^{k}_i \sigma^{k}_i) h^{k}_i \sigma^{k}_i
- \sum_{k=1}^{L-1} h^{k}_{i_k} \sigma^{k}_{i_k} \right\ra_{AB,\TT,\b} \ ,
\end{split}
\label{thermint}
\eeq
\end{widetext}
which can be computed as a function of temperature by means of the path sampling algorithm.
A numerical interpolation yields the final result:
\beq\label{thermintZ}
Z_{AB}(\TT,\beta) = Z_{AB}(\TT,\beta=0) \, \, e^{\int_0^\b d\b' \, U_{AB}(\TT,\b')} \ ,
\eeq
which we want to compute for $\b=1$.
In Figure \ref{Fig8} we explicitly show an example of the thermodynamic integration procedure. Specifically, 
we plot $U_{AB}(\TT,\b)$ as a function of $\beta$. This is the quantity that has to be
integrated, as in Eq.~(\ref{thermintZ}), to obtain $Z_{AB}(\TT,\beta=1)$.

Note that for $\TT=\io$, the function $Z_{AB}(\TT)$ must converge to
the equilibrium probability of $B$:
\beq
Z_{AB}(\TT\to \io,\b) = \la \chi_B(\bs) \ra_{eq} = 
\frac{\sum_{\bs} e^{-\b H(\bs)} \chi_B(\bs)}{\sum_{\bs} e^{-\b H(\bs)}} \ ,
\eeq
where $\la \bullet \ra_{eq}$ denotes the standard equilibrium thermodynamic average.
Then it is easy to show that
\beq
U_{AB}(\TT\to \io,\b) = - \la H \ra_{B} + \la H \ra_{eq} \ ,
\eeq
where 
\beq
\la H(\bs) \ra_{B} = \frac{ \sum_{\bs} H(\bs) e^{-\b H(\bs)} \chi_B(\bs) }{ \sum_{\bs} e^{-\b H(\bs)} \chi_B(\bs) } 
\eeq
is the average energy in the constrained Gibbs measure on state $B$.
Both $\la H \ra_{B}$ and $\la H \ra_{eq}$ can be quickly computed by a standard Monte Carlo simulation.
The results obtained from this simulation are reported as a dashed line in Fig.~\ref{Fig8}.

We see that for small enough $\b$ and a fixed $\TT$, the equilibration time is smaller
than $\TT$ so that the path Monte Carlo simulation result for $U_{AB}(\TT,\b)$ coincides with
$U_{AB}(\TT\to \io,\b)$. This is a crucial observation because it allows to avoid the path Monte
Carlo simulation at small $\b$ and large $\TT$, which is a difficult simulation since in this
regime the trajectories have a lot of jumps and the algorithm becomes very slow.

\subsubsection{The full $Z_{AB}(t)$ curves}

Using thermodynamic integration we can obtain $Z_{AB}(\TT)$ at $\beta=1$ for some values
of $\TT$. For each of these values of $\TT$, we can also estimate for free
the function $Z_{AB}(t)$
for all $t \leq \TT$ as discussed in section~\ref{sec:fullZt}. Namely, we compute
$\la \chi_B(t) \ra_{AB,\TT}$ in the path simulation at $\b=1$ and the chosen value of $\TT$, and we
use Eq.~(\ref{smallertime}).

\begin{figure}
\includegraphics[width =  \linewidth]{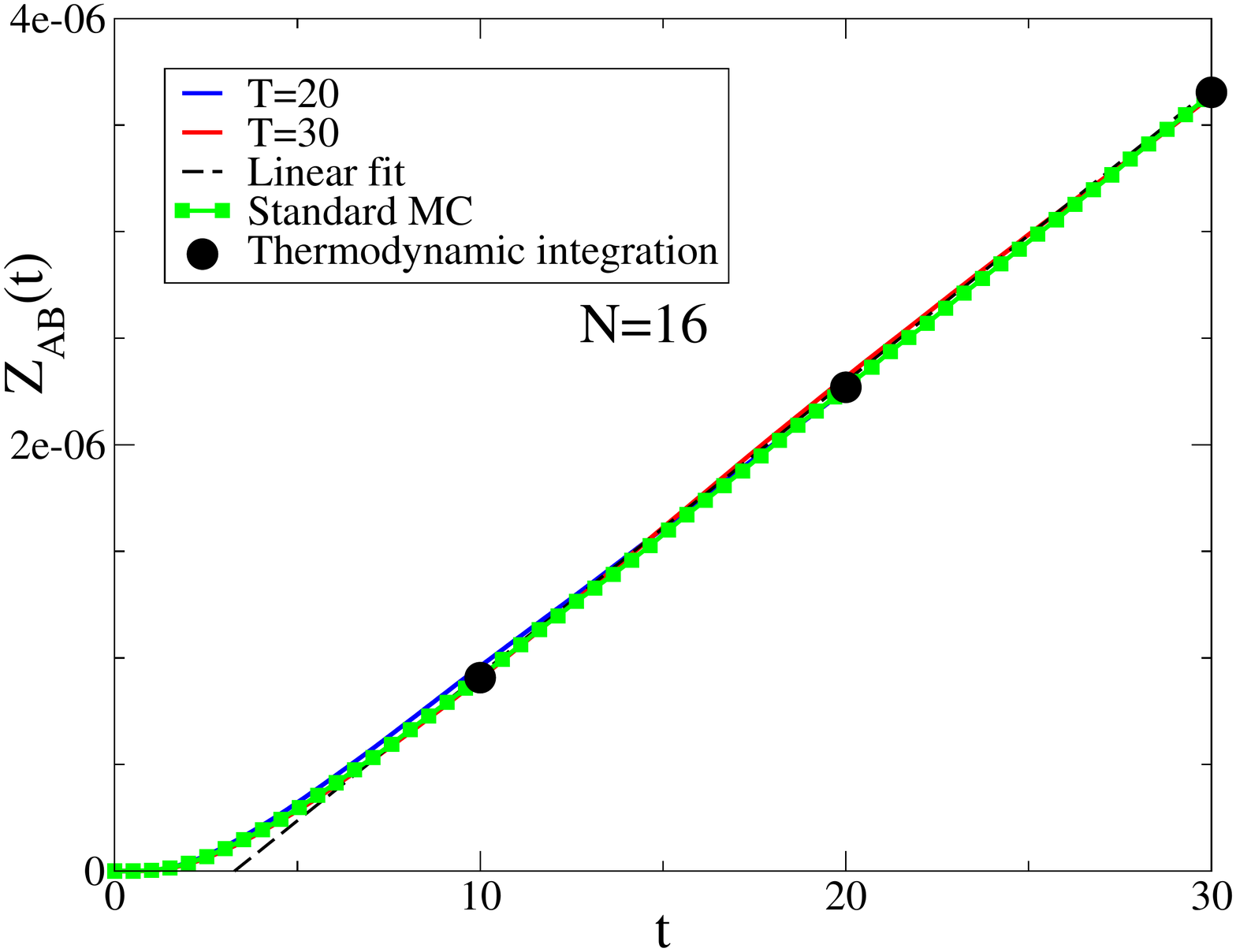}
\includegraphics[width =  \linewidth]{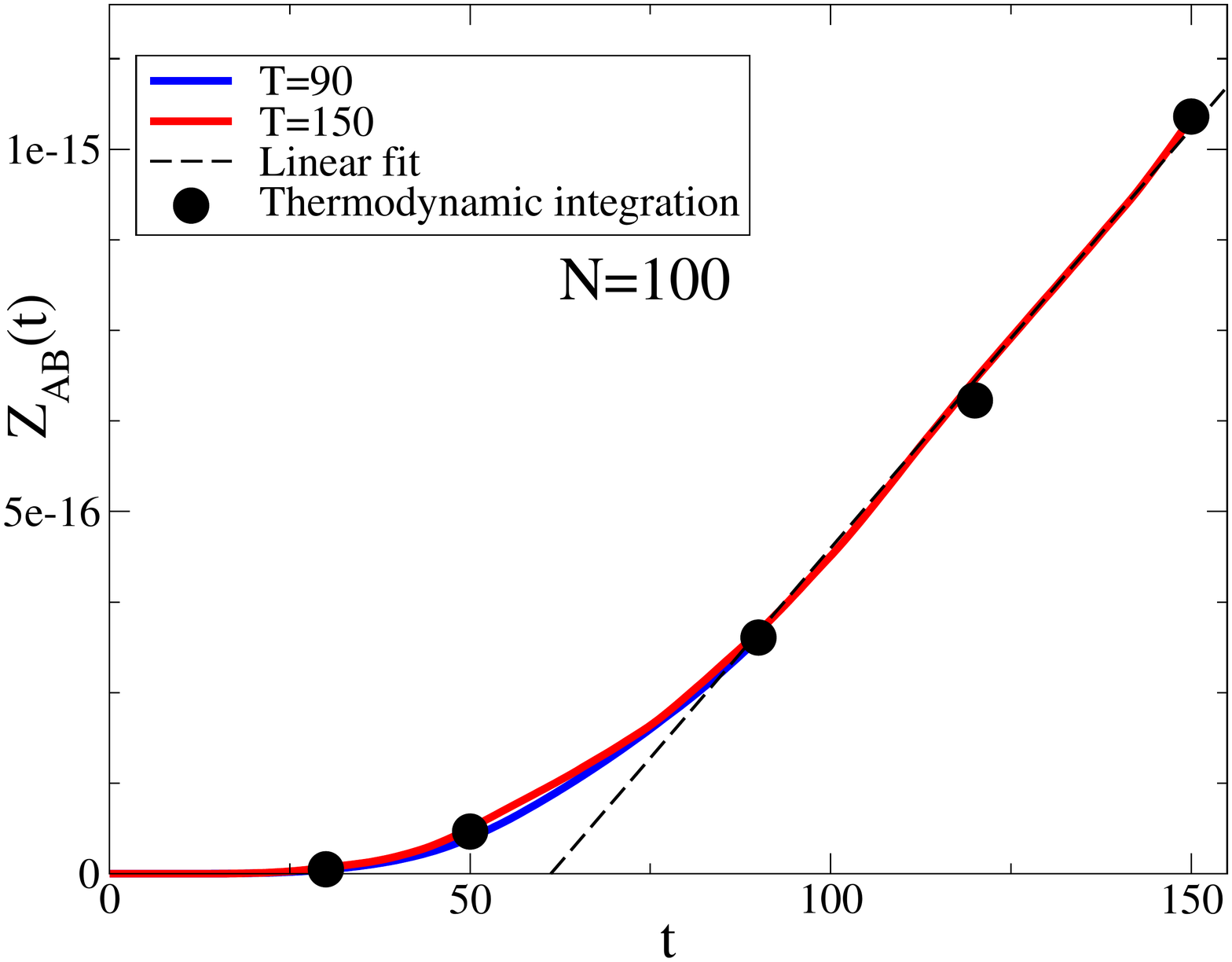}
\caption{Probability of transition $Z_{AB}(t)$ versus time in the 2D
  Ising model (here with $\beta=1$, $h_{A}=-3$, $h_{B}=1$).
Black dots represent the result of thermodynamic integration, the red and blue full lines
represent the result obtained using Eq.~(\ref{smallertime}) for two values of $\TT$.
After a transient time the function 
becomes linear, and straight dashed lines show the linear fit $Z_{AB}(t)=k_{A\to B} \times (t-\t_{\rm trans})$.
{\it Upper panel}: a small system with $N=16$ and $M^*=16$. For $N=16$ a direct comparison between our
  method and a traditional Monte Carlo simulation shows a perfect agreement.
{\it Lower panel}: a large system with $N=100$ and $M^*=56$. 
}
\label{Fig6}
\end{figure}

Fig.~\ref{Fig6} shows the full function $Z_{AB}(t)$, 
as calculated by this method, for $N=16$ and $N=100$.
Each panel of the figure shows a superimposition of two
curves, each corresponding to a different $\cal T$ (red and blue lines).
In addition, the values of $Z_{AB}(\TT)$ obtained by thermodynamic integration
are plotted as full black dots, for the available $\TT$.

For $N=16$, the transition rate is large enough ($k_{A \to B} \sim
10^{-7}$) so that we
can obtain a reliable result for the function $Z_{AB}(t)$ just by the ``naive'' Monte Carlo
approach, {\em i.e.} by running many standard faster-than-the-clock Monte Carlo simulations~\cite{krauth}
starting from the $A$ state and counting the fraction of them that is in state $B$ after a time $t$.
For $t = 10$, a fraction of $10^{-6}$ of such simulations is in state $B$, which means that in order to
have good statistics we only need to run $\sim 10^9$ independent
simulations for $N=16$ and $t=10$.
The result is reported with green full squares and show perfect agreement
with the path Monte Carlo simulations. 
On the other hand, for $N=100$ the rate is so small ($k_{A \to B} \sim 10^{-17}$) that obtaining a reliable
result by traditional Monte Carlo is completely impossible.

\begin{figure}
\includegraphics[width =  \linewidth]{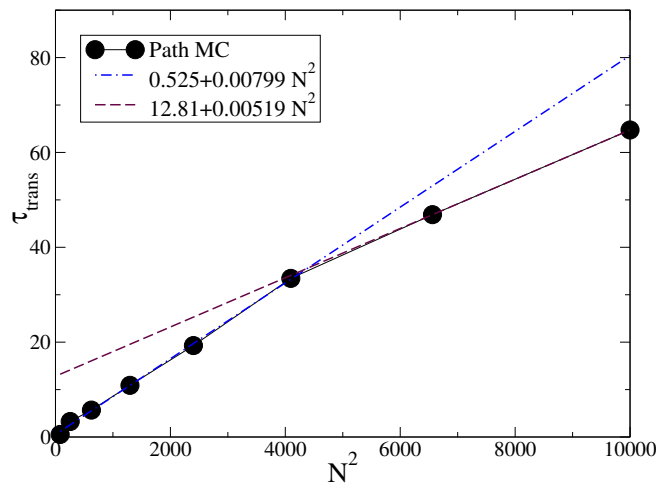}
\caption{Transient time $\tau_{\rm trans}$ needed to reach the linear scaling regime of
  $Z_{AB}(t)$, as a function of the square of the system's size $N^2$,
for $\beta=1$.}
\label{Fig7}
\end{figure}

Some minimum
time is required for the transition to occur (of the order of the time
necessary to relax to state $A$ or $B$), and thus the curve
$Z_{AB}(t)$ does not behave linearly with $t$ at very short times. After this initial transition
time however, the function becomes linear and can be fitted as
$Z_{AB}(t)=k_{A\to B} \times (t-\tau_{\rm trans})$ (dashed lines in Figure~\ref{Fig6}),
where $k_{A \to B}$ is the transition rate, and
$\tau_{\rm trans}$ is interpreted as the ``transient time''. 
In Fig.~\ref{FigureMAIN} we plot the logarithm of $k_{A\to B}$ as a
function of the system's linear dimension $L = \sqrt{N}$. In the same
figure we show the rates obtained from exact diagonalization at small sizes, when applicable.
The transition rate appear as an exponentially decaying function
of $\sqrt{N}$. Fig.~\ref{Fig7} suggests that $\tau_{\rm trans}$
depends quadratically on $N$. A crossover in the slope of the plots is observed around
$L = 8$; we will discuss this point below.

\begin{figure}
\includegraphics[width =  \linewidth]{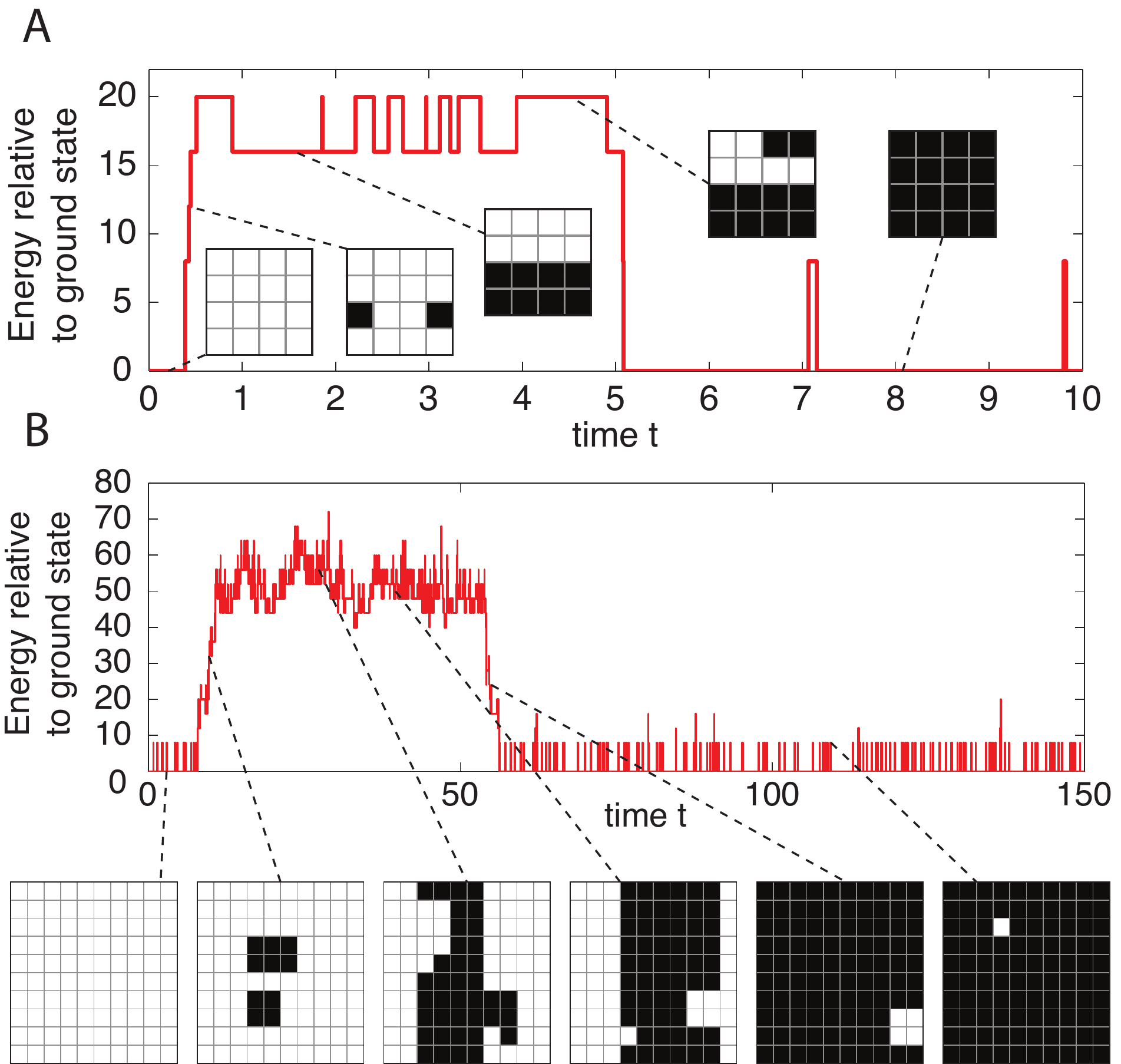}
\caption{Sample snapshots of transition paths in the 2D Ising model, for ({\bf A}) $N=16$ and ${\cal T}=10$, and ({\bf
    B}) $N=100$ and ${\cal T}=150$, at temperature $\beta=1$. White
  squares are down spins, and black squares are up spins.
}
\label{example}
\end{figure}

Our method allows us to inspect actual transition paths in detail. In
Fig.~\ref{example} we show two examples of snapshots transition paths, at
small ($N=16$) and large ($N=100$) sizes. The films of the full transition paths
are available as supplementary documents.
These examples show how the
system first creates a stripe of up spins in a background of down
spins. This stripe then progressively invades the lattice.

\subsubsection{Computer time}

To conclude this section we want to give an order of magnitude of the computer time that was needed to obtain
the above results. We want to stress, however, that our code was not particularly optimized for the model we investigated,
but it just corresponds to a plain implementation of the algorithm described above. We believe that its performances
might be improved by some smart optimizations, which are beyond the scope of this work.

The simulations 
were conducted on standard workstations, equipped with Intel Core i7 CPUs running at 2.80 GHz.
For the smallest systems, e.g. at $N=16$ and $\TT=10$, the calculation requires of the order of one day of
CPU time to obtain very accurate results. For the largest system we simulated ($N=100$ and $\TT=150$,
corresponding to the rightmost black point in lower panel of Fig.~\ref{Fig6}),
a single point of thermodynamic integration required a computational
time of the order of one month. 
The thermodynamic integration required running 12 independent values of temperature,
therefore obtaining $Z_{AB}(\TT)$ for $N=100$ and $\TT=150$ required a total of almost 1 year
of CPU time (which of course was possible in a much shorter time by using a small cluster of 48 cores).
The computational time for a given system scales as $N \TT$,
which is the ``system size''.
Overall, we believe that this is a good performance, because the value of the rate at $N=100$ is
extremely small ($k_{A\to B} \sim 10^{-17}$), so we are looking to really rare events.

Of course, as for any Monte Carlo simulation, 
the computational time depends crucially on the desired statistics.
Given the complexity of the procedure, we were unable to estimate error bars in a reliable way,
however we roughly estimate them to be of the order of symbol sizes in Fig.~\ref{Fig6}, which
we believe to be sufficient for the present purposes.

A final remark is that the computation of $U_{AB}(\TT,\b)$ turns out to be much easier than
that of $\la \chi_B(t) \ra_{AB,\TT}$ on the same state point. This is related to the following
observation. Typical configurations of the paths are given in Fig.~\ref{example}, and they are
characterized by periods of inactivity (in which all spins are up or down) separated by
the barrier crossing period, where the energy is above the ground state. The position of the latter
period fluctuates uniformly and slowly during the path Monte Carlo simulation. Remarkably, the value
of $U_{AB}(\TT,\b)$ is independent of the time location of the barrier crossing, therefore one does
not need to accumulate much statistics on the slow fluctuations of the latter to have a reliable result
on this quantity.
On the contrary, the calculation of $\la \chi_B(t) \ra_{AB,\TT}$ clearly requires a perfect sampling
of the fluctuations of the barrier crossing point. Achieving this seems much more difficult and for
this reason this quantity is typically much more noisy. 
For this reason we found that the results of thermodynamic integration (black dots in figure~\ref{Fig6})
were typically much more reliable that the ones obtained through $\la \chi_B(t) \ra_{AB,\TT}$ (red and blue
lines in Fig.~\ref{Fig6}).

\subsection{Interpretation}\label{Asymscaling}

\subsubsection{Transition state and surface tension}

In the limit of large system sizes, simple arguments allow us to
write the scaling of the transition rate in terms of the surface tension,
$\Sigma$. If we imagine starting from a homogenous system in which all
spins are aligned, the transition time depends on the probability of the
initial nucleation event of the first stripe of spins of opposite
sign. The escape time is then simply proportional to the exponential
of surface tension $\Sigma$ of that stripe, times its surface
$2\sqrt{N}$ ($\sqrt{N}$ is the length or linear dimension of the system, and the stripe
has two interfaces)~\cite{martinelli}:
\beq
k_{A\to B} \sim \exp(-2 \Sigma \sqrt{N} ).
\label{gap}
\eeq

\begin{figure}
\includegraphics[width = .8 \linewidth]{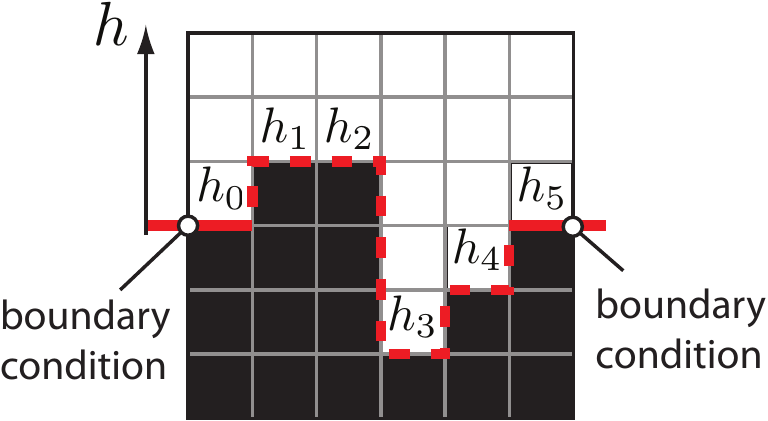}
\caption{
In the approximated description of~\cite{zittarz}, the height of the
interface is assumed to be 
a single-value function, or in other words the overhangs are ignored.
}
\label{model}
\end{figure}

The surface tension in a 2D Ising model has been calculated exactly in the
thermodynamic limit by Onsager, and its value is $\Sigma = 2 \beta J + \log \tanh
\beta J$~\cite{onsager, gallavotti, binder, martinelli}. However, a simple 
model~\cite{zittarz} allows to obtain an approximated expression for
the surface tension also at finite size. In this simplified model, illustrated in 
Fig.~\ref{model}, an interface between a plus and minus region is described
by a single-valued function $h_i$, neglecting overhangs. At the left and right
boundary the interface is supposed to be in $h_0 = h_{L+1} =0$. The energy
associated with the interface is
\beq
H_I = 2 L + 2 \sum_{i=1}^{L+1} | h_i - h_{i-1} | \ ,
\eeq
and its partition function is therefore
\beq
 Z_I = e^{-2\b L} \sum_{h_1\cdots h_{L}}  e^{-2 \b  \sum_{i=1}^{L+1} | h_i - h_{i-1} | } \ ,
\eeq
which is easily computed by Fourier transform:
\beq\label{ZI}
Z_I = e^{-2 \b L} \int_{- \pi}^\pi \frac{dk}{2\pi} \left( \frac{ e^{4 \b}-1}{e^{4\b} +1 - 2 e^{2\b} \cos(k)} \right)^{L+1} \ .
\eeq
The corresponding surface tension is $\Si = -L^{-1} \log Z_I$.

The result is that at small $L$ and large enough $\b$, 
the partition function is dominated by the configuration $h_i =0$
and tends to be rectilinear, thereby losing the
benefit of the entropic contribution to the surface tension. Such a
rectilinear boundary leads
to $\Sigma=2\beta J$.
On the contrary, for large $L$ the integral in Eq.~(\ref{ZI}) can be evaluated by a saddle-point
and gives the exact result of Onsager, including the entropic contribution.
The crossover between these two regimes happens at a length scale that is extremely small
close to $\b_c$, and grows with decreasing temperature, diverging at $\b \to \io$. From Eq.~(\ref{ZI}),
we find that at $\b=1$ the crossover indeed happens around $L=10$.

These results are consistent with the data we reported in Figure \ref{FigureMAIN}.
Indeed, the slope at small $L$ is consistent with a
higher surface tension $\Si=2$ (in units of $J$, and for $\b=1$). 
As $L$ increases, the slope asymptotically
approaches the correct value $\Si = 1.7276$ given by the Onsager formula.

\begin{figure}
\includegraphics[width =  \linewidth]{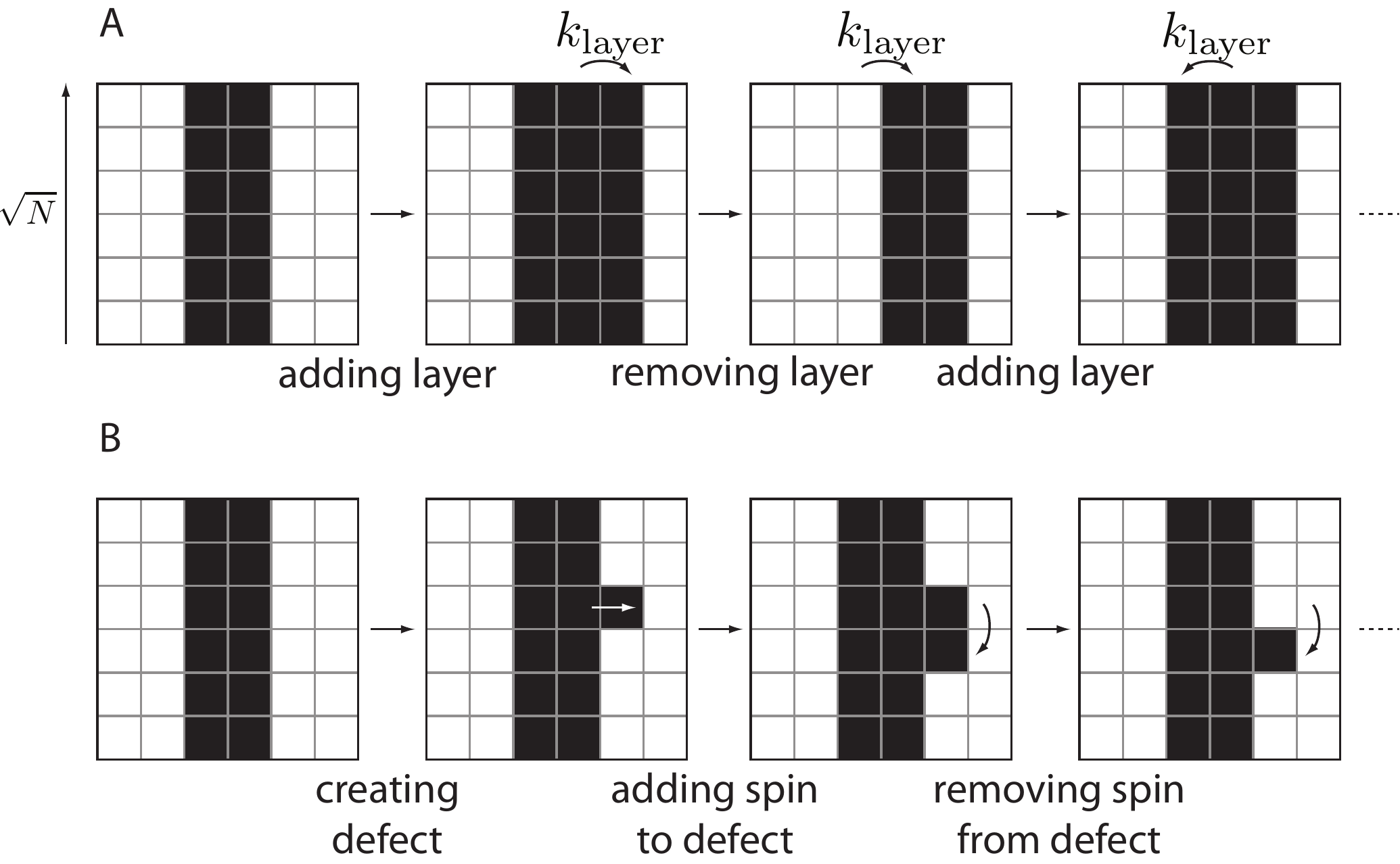}
\caption{Zero-temperature relaxation from a stripe to the
  final state, viewed as a double random walk. Black boxes are up
  spins, and white boxes and down spins. {\bf A.} Adding and removing layers to the stripe is
  can be described by a random walk with stepping rate $k_{\rm
    layer}$. The total time for the stripe invading the system scales
  like $k_{\rm
    layer}^{-1} (\sqrt{N})^2$. {\bf B.} A new layer is added/removed by first creating a
  defect on the boundary, then propagating it across the system's
  length. This propagation is also governed by random walk, and takes time $k_{\rm
    layer}^{-1} \sim (\sqrt{N})^2 $.}
\label{Figstripe}
\end{figure}

\subsubsection{Transient time}

We find that the transient time grows linearly with $N^2$. Let
  us try to interpret this result. The
transient time may be interpreted as the minimum time for the
transition to occur. This time is at least as long as the time the system
takes to relax to state $B$ after starting at the top the barrier
between $A$ and $B$. In our case, the top of the barrier corresponds
to configurations where a stripe of up spins has nucleated across the
system's length in a background of down spins. Let us reason at
low enough temperature, where the stripe is almost perfectly rectilinear and
where defects are extremely rare. We expect our reasoning to hold for
arbitrary temperatures.
In order to invade the lattice entirely, the stripe needs to
thicken by adding new layers of up spins on either of its
sides. Likewise, the stripe may thin out through the removal of 
spin layers. The thickness of the stripe therefore undergoes an
unbiased random walk (Fig.~\ref{Figstripe}A).
If the rate of adding or removing layers to the stripe is
$k_{\rm layer}$, then the expected time for the stripe to invade the
system is the time it takes for the random walk to reach the system's
length, $\sqrt{N}$. This time scales as $\sim k_{\rm layer}^{-1}
(\sqrt{N})^2=k_{\rm layer}N$. The rate $k_{\rm layer}$ itself can be estimated
in a similar manner (Fig.~\ref{Figstripe}B). A new layer can be added when an up spin appears
sticking out from one of the two stripe's boundaries, incurring a $4J$ energy cost. Once
such a defect has been created, adding or removing up spins on the same layer
contiguously to the defect---to make it bigger or smaller---has no energy cost. The length of the defect is
therefore governed by a random walk, which ends when the defect
disappears or creates a new layer of up spins by reaching the
system's length $\sqrt{N}$. Layer removal occurs is
the exact same way, and therefore $k_{\rm layer}^{-1}\sim
(\sqrt{N})^2=N$. In summary, we expect the transient time to scale
with $N$ as $k_{\rm layer}^{-1}N\sim N^2$, in agreement with our
results. 
It seems that the prefactor of this scaling depends on the surface tension,
since we observe that it changes around the same length scale as
in Fig.~\ref{FigureMAIN}.

\section{Conclusions}\label{sec:conclu}

We have presented a method for efficiently sampling transition
trajectories of discrete systems. The method is general and applicable
to systems in and out of equilibrium. The method scales well with
system size and we are able to effectively sample trajectories in
systems for which asymptotic scaling holds. We emphasize two main
advantages of the method presented in this paper. Firstly, the method
does not require detailed balance to hold and is therefore applicable
to all nonequilibrium systems.
Secondly, the method scales reasonably with the number of variables in
the system, allowing one to calculate the escape rates for otherwise
prohibitively large problems.
Combined with a thermodynamic integration procedure,
it gives detailed information on the typical transition paths,
the full time-dependent transition probability $Z_{AB}(t)$,
the transition rate $k_{A\to B}$, and 
the transient time $\t_{\rm trans}$, which is related to the minimal
barrier crossing time.
We tested our method on the equilibrium example of a 2D Ising model
with periodic boundary conditions, where we achieved excellent
agreement between the results of the path sampling method to exact
matrix diagonalization and predictions for asymptotic scaling.

Our method is based on the classical path sampling
method of Dellago {\em et al.} for continuous variables \cite{chandler}, which is
based on the
idea of avoiding performing a detailed Monte Carlo simulation on the
variables of the system, but instead to propose a Monte Carlo
algorithm on the paths themselves, and on the use of thermodynamic
integration to compute the transition rate.

Our new implementation of the path sampling takes advantage of
the discrete and many-body nature of the system.
It allows us to consider the trajectories for each variable separately and
modify them while keeping the rest of the system fixed. As a result we can
\textit{draw} a whole new trajectory for the chosen spin, instead of just
modifying it locally. We expect this procedure to be more efficient in sampling the 
space of paths. We also solved a technical issue that is specific to many-body
systems, namely the fact that the thermodynamic integration on the final state
proposed in~\cite{chandler} fails because a first order phase transition is met on the
integration path. We performed instead a thermodynamic integration in temperature;
thermodynamic quantities are smooth on this path, allowing for an accurate computation.
We expect this to be a generic phenomenon for many-body systems.

A direct comparison with the performances of 
other methods is not straightforward. Some of these methods
have been applied on the 2D Ising model~\cite{shearising, venturoli} but the nucleation 
problem studied there was different (nucleation in presence of an external magnetic field).
Also, all these methods are complex enough that the actual performances depend a lot
on the implementation and the details of the problem under investigation.
We believe that the important point is that the present method scales linearly with the ``size''
of the path system, $N \TT$ (note however that the time $\TT$ typically grows polynomially with 
system size, e.g. $\TT \propto N^2$ in this case, meaning that the overall computational time
is expected to scale polynomially in $N$ with some exponent larger than 1). Thanks to this,
we believe that the method gives an interesting way to study discrete many-body problems
and obtain complementary information to other techniques.

The method can be straightforwardly
applied to study the dynamics of disordered spin systems,  chemical
reactions and gene regulatory systems.  In these last two classes of
problems, one needs to consider the numbers of molecules that take
part in the reactions---a number that is in principle infinite, but
usually bounded in practice. The method presented here is still
applicable to such problems. Although the purpose of the present paper is to
introduce the method in a clear way and convincingly show agreement
with well known results on an
equilibrium example, future work should focus on non-equilibrium
applications.

A very interesting property of this method is that it allows us to examine typical
sample trajectories and get a detailed picture of the transition, with
{\em e.g.}
the detailed shape and dynamics of the critical nucleus in the 2D
Ising model.
In particular,
we get complete access to the time-dependent cumulative
transition probability $Z_{AB}(t)$, which tells us the probability that
the system undergoes a transition {\it even for very short times when the
process
is not yet Poissonian}---{\em i.e.} for times shorter than the typical
relaxation time. In the 2D Ising model, we estimated this transient
time and discussed its scaling with the system's size. The estimation
of $Z_{AB}(t)$ at short times
might also be important for some biological applications where
one deals with large microbial populations. In such
problems, the rarity of transition events in gene-regulatory or
biochemical networks is compensated by very large size of
populations, which makes that nominally rare events occur quite often
at the population level. In this context, $Z_{AB}(t)$ may for example be
interpreted as the fraction of individuals that make a potentially
life-saving transition within some finite time $t$ after the
introduction of a stress. Our method provides the tools to estimate
such tiny fractions in models of biochemical networks \cite{suel, vano}.

\acknowledgments

We would like to warmly thank Patrick Charbonneau, Zoran Ristivojevic, and Guilhem Semerjian for several crucial discussions. 
We acknowledge the support of the Projet Incitatif de Recherche 2011 grant from the \'Ecole normale sup\'erieure.

\appendix

\section{Detailed calculations for the mean field
model}\label{AppendixMF}

\subsection{Master equation}

Because the Hamiltonian depends only on $M$, 
it follows that at any time $t$,
$p_t(\bs)$ depends only on $M$ too (provided this is true at
$t=0$). Therefore we can write:
\beq
p_t(\bs) = p_t(M) \binom{N}{(N+M)/2}^{-1}.
\eeq 
Injecting the above equation into Eqn.~(\ref{mastereqn}), and using
the relation
\beq
\frac{N+M \sigma}2 \frac{\binom{N}{(N+M)/2}}{\binom{N}{(N+M-2\sigma)/2}}
= \frac{N- (M-2\sigma) \sigma}2,
\eeq
it is easy to show that
\beq\label{MFmaster}
\begin{split}
\partial_t p_t(M) &= w_+(M-2) p_t(M-2) + w_-(M+2) p_t(M+2) \\
& - [w_-(M) + w_+(M)] p_t(M) = \LL p_t,
\end{split}\eeq
with
\beq\begin{split}
w_+(M) &= \frac{N-M}2 w \left[-2 \frac{M+1}N \right]=\frac{N-M}2 e^{\beta (M+1)/N}, \\
w_-(M) &= \frac{N+M}2 w \left[2 \frac{M-1}N \right]=\frac{N+M}2 e^{-\beta(M-1)/N},
\end{split}\eeq
which has the form of a one-dimensional birth-death process~\cite[Section 7.1]{gardiner}.

\subsection{Mean first-passage time}

We then use the results of~\cite[Section 7.4]{gardiner} for discrete, one-dimensional birth-death processes
in order to compute the mean
first passage time in $M_{end}$
of a system that starts in $M_{start} < M_{end}$ at time $t=0$ (hence we are taking the negative $M$ state
as the initial state, and the positive $M$ state as the final state).
Obviously the system is confined by a reflecting barrier in $M = -N$.
Using this and~\cite[Eq.~(7.4.12)]{gardiner}, we get
(in the following sums, capital letter $K,L,M$ 
denote magnetizations and therefore increase in steps of 2 units)
\beq\label{MF_MFPT}
T(M_{start} \to  M_{end}) = \sum_{K = M_{start}}^{M_{end}} \phi(K) \sum_{L=-N}^K \frac{1}{\phi(L) w_+(L)} \ ,
\eeq
with
\beq
\phi(M) = \prod_{K=-N+2}^M \frac{w_-(K)}{w_+(K)}.
\eeq
The latter expression can be computed numerically for finite $N$, in a time growing only polynomially in $N$.

\subsection{Large $N$ limit}
We want to study the large $N$ asymptotic behaviour of Eq.~(\ref{MF_MFPT}).
We note that using Eq.~(\ref{wrate}) we have
$w_-(M)/w_+(M) = \exp[2\b f'(M/N)]$
where $f'(m) = (1/2) \left( \log [(1+m)/(1-m)]-2 \b m \right)$ is the derivative of the free energy
in Eq.~(\ref{MFstaticf}). Therefore,
\beq\begin{split}
\phi(m N) &= \exp\left[\sum_{K=-N+2}^M 2 \b f(K/N)\right] \\
&=
\exp\left\{N \b [ f(m) - f(-1) ] + \D(m) \right\}
\end{split}
\eeq
with
\beq\begin{split}
\D(m) = &\b \sum_{K=-N+2}^M \left[ 2 f'\left(\frac{K}{N}\right) - N
  f\left(\frac{K}{N}\right) \right.\\
&+ \left. N f\left(\frac{K-2}{N}\right) \right]
\end{split}\eeq
To estimate the correction $\D(m)$, we need to separate the free energy in Eq.~(\ref{MFstaticf}) in two terms:
\beq\label{MFsplitting}
\b f(m) = \b f_{\rm reg}(m) +\frac{1+m}2 \log \frac{1+m}2
\eeq
where the second term is singular at $m=-1$. Likewise, we separate
$\D(m)=\D^{\rm reg}(m)+\D^{\rm sing}(m)$ into a regular and a singular
term.
For the first term, we can use that
$\frac{2}{N} f'_{\rm reg}\left(\frac{K}{N}\right) - f_{\rm reg}\left(\frac{K}{N}\right) + f_{\rm reg}\left(\frac{K-2}{N}\right) \sim 
\frac12 \left(\frac{2}{N}\right)^2 f''_{\rm reg}\left(\frac{K}{N}\right)$, and therefore
\beq\begin{split}
\D^{\rm reg}(m) &\sim \b \sum_{K=-N+2}^M \frac{2}{N} f''_{\rm reg}\left(\frac{K}{N}\right)
 \sim \b \int_{-1}^m dk f''_{\rm reg}(k) \\
& = \b [ f'_{\rm reg}(m) - f'_{\rm reg}(-1) ] 
\end{split}\eeq
For the singular term, we can use the explicit form of the second term in (\ref{MFsplitting})
to write
\beq\begin{split}
\D^{sing}(m) &= \sum_{K=-N+2}^M \left( 1 + \frac{K+N-2}2 \log \frac{K+N-2}{K+N} \right) \\
& = \sum_{K=0}^{M+N-2} \left( 1 - \frac{K}2 \log \frac{K+2}{K} \right) \\
&\sim \frac12 \log[\pi N (m+1)] + O(1/N)
\end{split}\eeq
where the last line can be obtained by recognizing that the sum can be written as a convergent
part plus a divergent sum which is the harmonic number, and then using the asymptotic expression
of the latter. We get the final result
\beq
\phi(mN) = \sqrt{\pi N (m+1)} e^{\b N [f(m)-f(-1)] + \D^{\rm reg}(m) }
\eeq

\begin{widetext}

Next we evaluate the sum
\beq\begin{split}
&\phi(M)  \sum_{K=-N}^M \frac{1}{\phi(K) w_+(K)} 
\sim  \sqrt{m+1} e^{\b N f(m) + \D^{\rm reg}(m) }     \int_{-1}^m dk  \frac{e^{-\b N f(k) - \D^{\rm reg}(k) }}
{\sqrt{k+1} (1-k)e^{\b k}} \\
&\sim \sqrt{ \frac{2 \pi}{N \b f''(m^*)} }  \sqrt{ \frac{1+m}{1-m^*} }
\frac{e^{\b N [f(m)-f(m^*)] + \D^{\rm reg}(m)- \D^{\rm reg}(-m^*)}}
{(1-m^*)e^{-\b m^*}}
\end{split}\eeq
where the second line is obtained via the saddle point method. The saddle point is at $k=-m^*$,
and we assumed that $m > -m^*$ which
is the case that will be relevant in the following. We also used the symmetry $f(m)=f(-m)$.

Finally, recalling Eq.~(\ref{MF_MFPT}):
\beq
T(M_{start} \to M_{end}) \sim \frac{N}2 \int_{m_{start}}^{m_{end}} dk
  \sqrt{ \frac{2 \pi}{N \b f''(m^*)} }  \sqrt{ \frac{1+k}{1-m^*} }
\frac{e^{\b N [f(k)-f(m^*)] + \D^{\rm reg}(k)- \D^{\rm reg}(-m^*)}}
{(1-m^*)e^{-\b m^*}}
\eeq
Assuming that $m_{end}>0$ and $m_{start} \in [-m^*,0]$, we can again evaluate the integral by a saddle point,
the saddle point being in $k=0$ in this case:
\beq
T(M_{start} \to M_{end}) \sim \frac{N}2 
\frac{2 \pi}{N \b} \sqrt{ \frac{1}{ f''(m^*) | f''(0)|} }  \sqrt{ \frac{1+k}{1-m^*} }
\frac{e^{\b N [f(0)-f(m^*)] + \D^{\rm reg}(0)- \D^{\rm reg}(-m^*)}}
{(1-m^*)e^{-\b m^*}}
\eeq
Simplifying this expression leads to the final result 
\beq
\text{MFPT}_{A\to B} =
\frac{\pi}{\b} \sqrt{\frac{1}{[1-\b(1-(m^*)^2)](\b-1)}}e^{\b N [f(0) -
  f(m^*) ]}.
\eeq
as reported in Eq.~(\ref{MF_MFPT_largeN}).

We notice that the above result is independent of $M_{start}, M_{end}$ provided they scale proportionally
to $N$. Alternatively,
one can consider a scaling regime where $M_{end} = \sqrt{N} y$ and $M_{start} = \sqrt{N} x$,
in which case
\beq\label{MF_MFPT_scaling}
\frac{T(\sqrt{N} x \to \sqrt{N} y)}{T(-\io \to \sqrt{N} y)} = 
\frac{\int_{x \sqrt{N}}^{y \sqrt{N}} dk e^{\b N f''(0) k^2}}
{\int_{-\io}^{y \sqrt{N}} dk e^{\b N f''(0) k^2}}
= \frac{\text{erf}(\k y) - \text{erf}(\k x)}{\text{erf}(\k y) -1}
\eeq
with $\k = \sqrt{-\b f''(0)}$. This scaling regime is the one where the mean first passage time
depends on the initial and final points.

\end{widetext}

\subsection{Calculation of $Z_{AB}(t)$}

For the mean-field model, the function $Z_{AB}(t)$ can be defined as follows:
\beq
Z_{AB}(t) = \sum_{M,M'} \chi_B(M) \big( e^{\LL t} \big)_{M,M'} p_A(M')
\eeq
where the operator $\LL$ is defined in Eq.~(\ref{MFmaster}), $p_A(M)$ is an
initial probability distribution which is assumed to be centered on state $A$,
and $\chi_B(M)$ is the indicator function of state $B$, i.e. it is one when the
system is in state $B$ and zero otherwise.

This quantity
can be easily computed. Recalling that the invariant distribution is 
\beq
p_{eq}(M) = \binom{N}{\frac{N+M}2} e^{-\b \frac{M^2}{2N}} \ ,
\eeq
we can define a symmetric matrix
\beq\begin{split}
H_{M,M'} & = \sqrt{\frac{p_{eq}(M')}{p_{eq}(M)}} \LL_{M,M'} \\
& = \frac12 \sqrt{(N-M+2)(N+M)} \, \d_{M',M-2} \\
& + \frac12 \sqrt{(N+M+2)(N-M)} \, \d_{M',M+2} \\
 -e^{\b/N}  & \big[ N  \cosh(\b M/N) - M \sinh(\b M/N) \big] \d_{M',M}
\end{split}\eeq
which can be easily diagonalized helding eigenvalues $v_n(M)$ and 
(negative) eigenvectors $\l_n$, and 
\beq\label{ZABMF}
\begin{split}
Z_{AB}(t) & = \sum_{n} e^{\l_n t} 
 \left(\sum_{M} \chi_B(M) \sqrt{p_{eq}(M)} v_n(M) \right) \\
& \times \left(\sum_{M}  \frac{v_n(M)}{\sqrt{p_{eq}(M)}} p_A(M) \right)
\end{split}\eeq
To produce the plots of Fig.~\ref{Figure_MF_Zt}, 
we chose as a definition of the states $A$ and $B$ the following functions:
\beq\begin{split}
&p_A(M) = e^{h_A M}/[2 \cosh(h_A)]^N \ , \\
&\chi_B(M) = e^{-h_B (M^* - M) \theta(M^* - M)} \ ,
\end{split}\eeq
with $h_A = -3$, $h_B = 1$ and $M^* = 2N/3$. 
However, the shape of $Z_{AB}(t)$ is largely independent of the 
details of these definitions.

\end{document}